# Nonlinear Adaptive Cruise Control of Vehicular Platoons

**Iasson Karafyllis[*], Dionysis Theodosis[**] and Markos Papageorgiou[**]**

[*]Dept. of Mathematics, National Technical University of Athens,
Zografou Campus, 15780, Athens, Greece,
emails: iasonkar@central.ntua.gr , iasonkaraf@gmail.com

[**] Dynamic Systems and Simulation Laboratory,
Technical University of Crete, Chania, 73100, Greece
(emails: dtheodosis@dssl.tuc.gr , markos@dssl.tuc.gr)

**Abstract**

The paper deals with the design of nonlinear adaptive cruise controllers for vehicular platoons operating on an open road or a ring-road. The constructed feedback controllers are nonlinear functions of the distance between successive vehicles and their speeds. It is shown that the proposed novel controllers guarantee safety (collision avoidance) and bounded vehicle speeds by explicitly characterizing the set of allowable inputs. Moreover, we guarantee global asymptotic stability of the platoon to a desired configuration as well as string stability. Certain macroscopic properties are also investigated. The efficiency of the nonlinear adaptive cruise controllers is demonstrated by means of a numerical example.

## 1. Introduction

The extension of the standard cruise control system is the adaptive cruise control (ACC) system, which has dual operation: In speed control mode, ACC maintains constant desired speed, as conventional cruise control; but if there is a slower vehicle ahead, ACC switches to spacing control and uses on-board sensors to reach a desired spacing from the preceding vehicle. This may lead to multiple ACC-equipped vehicles to form a platoon with tight vehicle spacing, which may increase safety and traffic flow capacity and reduce fuel consumption.

A large variety of spacing policies and controllers for ACC vehicles and platoons have appeared, see [1], [3], [4], [5], [6], [9], [18], [24], [14], [25], [22], [20], [23], [30], [33]. The most common policies considered in the related literature are the constant spacing policy [28], where the distance between successive vehicles remains constant at all speeds; and the Constant-Time Gap (CTG) policy [14], where the spacing varies linearly with speed. To evaluate a spacing policy and its associated controller, the following criteria were proposed, see [23]: (i) individual vehicle stability, which characterizes the convergence towards a desired equilibrium; (ii) string stability, which focuses on the dissipation of small perturbations along a string of vehicles ([6], [19], [27], [29]); and (iii) traffic flow stability which deals with the evolution of density when all vehicle use the same spacing policy ([23], [24], [26]).

The notion of string stability has been widely studied and several definitions have appeared in the literature, see [6], [19], [22], [29], [27], [32]. A detailed overview of the various string stability definitions and their properties can be found in [10], [19]. To distinguish the ambiguity over the different definitions used in the literature, a novel definition was proposed in [19] for both linear and nonlinear systems based on $L_p$ stability, which encompasses the upstream disturbance attenuation, the external input of the leading vehicle, as well as perturbations on initial conditions.

In addition to the stability properties above, a desired requirement on the vehicle platoon is the safe operation, which forbids collisions between vehicles, negative speeds and speeds exceeding speed limits. While string stability ensures that disturbances in position, speed or acceleration do not accentuate while propagating along the platoon, it does not guarantee collision avoidance between vehicles in the platoon, see [8]. Indeed, the majority of spacing policies and ACC controllers focus on stability and string stability properties ([12], [18], [33], [29], [34], [35]), which, however, may result in negative spacing error and negative speeds. On the other hand, approaches considering safety can be found in [1], [5], but they do not formally study string stability or operation on a ring-road; and also in [11], [13], [30], which mainly deal with boundedness of spacing error rather than convergence to a desired value. In [17], different control configurations and conditions for a CTG policy are derived that guarantee string stability and collision avoidance when the platoon is initiated from an equilibrium position with zero speed and sufficiently large initial spacing between vehicles. Safety criteria were also presented in [3], where collisions are avoided whenever the platoon does not exceed a given relative speed threshold regardless of the behavior of the leader. In the companion paper [4], single-lane maneuvers are studied that always imply safety. The proposed controller in [3], [4] can guarantee stability and string stability if the leading vehicle's acceleration and speed is known or estimated by all following vehicles.

It is clear from the above that a methodology that simultaneously guarantees safety, stability, string stability under predecessor-follower (i.e. autonomous) control architecture is missing in the literature. In this paper, we present conditions which guarantee safety, stability and $L_p$ string stability of a vehicular platoon using nonlinear adaptive cruise controllers. Due to the different conditions applying for an open road and a ring-road, we consider both cases separately and show that the proposed nonlinear controller has the following features:
1. It provides safe platoon operation without collisions, negative speeds or speeds exceeding speed limits.
2. It guarantees global asymptotic stability of the spacing/speed equilibrium for a platoon on an open road and global exponential stability for the case of a ring-road.
3. It guarantees $L_p$ string stability for the platoon.

Moreover, we explicitly characterize the set of feasible initial states for safe operation in terms of collision avoidance and bounded vehicle speeds, as well as the class of inputs (maneuvers of the leading vehicle) that can be allowed for the safe operation of the platoon. Finally, certain macroscopic properties related to traffic flow stability, design of the fundamental diagram, and the reduction of the microscopic model to the standard Lighthill-Witham-Richards (LWR) model [16], [21] are studied. The proofs of all results follow by explicit construction of Lyapunov functions and Barrier functions, [36]. The main difficulty that arises is due to the fact that the control systems studied in the paper do not evolve in a finite dimensional linear space but rather on specific open or closed sets.

The structure of the paper is as follows. Section 2 of the paper is devoted to the presentation of the properties of adaptive cruise controllers, such as safety criteria and appropriate stability notions. To facilitate the motivation for the use of nonlinear controllers, simulation scenarios are also presented in Section 2 using the standard CTG controller (see [20]), which demonstrate that certain safety criteria may fail. A general form of a nonlinear adaptive cruise controller is provided in Section 3 together with sufficient conditions for the safe operation of a platoon of vehicles both on an open road and on a ring-road. Section 4 provides results for the $L_p$ string stability of the proposed adaptive cruise controller. In Section 5, it is shown that the sufficient conditions for string stability and the existence of a fundamental diagram also guarantee global asymptotic stability of the unique equilibrium point of a platoon operating in an open road and global exponential stability for the case of a ring-road. Numerical examples are presented in Section 6 to demonstrate the efficiency of the proposed nonlinear adaptive cruise controller. All proofs of the main results are provided in Section 7. Finally, concluding remarks are given in Section 8.

**Notation.** Throughout this paper, we adopt the following notation.

* $\Re_+ := [0, +\infty)$ denotes the set of non-negative real numbers.
* By $|x|$ we denote both the Euclidean norm of a vector $x \in \Re^n$ and the absolute value of a scalar $x \in \Re$.
* By $K$ we denote the class of strictly increasing $C^0$ functions $a : \Re_+ \to \Re_+$ with $a(0) = 0$. By $K_\infty$ we denote the class of strictly increasing $C^0$ functions $a : \Re_+ \to \Re_+$ with $a(0) = 0$ and $\lim_{s \to +\infty} a(s) = +\infty$. By $KL$ we denote the set of all continuous functions $\sigma : \Re_+ \times \Re_+ \to \Re_+$ with the properties: (i) for each $t \geq 0$ the mapping $\sigma(\cdot, t)$ is of class $K$; (ii) for each $s \geq 0$, the mapping $\sigma(s, \cdot)$ is non-increasing with $\lim_{s \to +\infty} \sigma(s, t) = 0$.
* By $C^0(A, \Omega)$, we denote the class of continuous functions on $A \subseteq \Re^n$, which take values in $\Omega \subseteq \Re^m$. By $C^k(A; \Omega)$, where $k \geq 1$ is an integer, we denote the class of functions on $A \subseteq \Re^n$ with continuous derivatives of order $k$, which take values in $\Omega \subseteq \Re^m$. When $\Omega = \Re$ the we write $C^0(A)$ or $C^k(A)$.
* By $L^p$ with $p \geq 1$ we denote the equivalence class of measurable functions $f : \Re_+ \to \Re^n$ for which $\|f\|_{[0,t],p} = \left( \int_0^t |f(x)|^p dx \right)^{1/p} < +\infty$. $L^\infty$ denotes the equivalence class of measurable functions $f : \Re_+ \to \Re^n$ for which $\|f\|_{[0,t],\infty} = ess\sup_{x \in [0,t)} (|f(x)|) < +\infty$.
* For a set $S \subseteq \Re^n$, $\bar{S}$ denotes the closure of $S$.

## 2. Motivation

A commonly used model for vehicle dynamics in vehicular platoons consists of the following ODEs:

$$\begin{aligned} \dot{s}_i &= v_{i-1} - v_i \\ \dot{v}_i &= u_i \end{aligned}, \quad i = 1, ..., n \quad (2.1)$$

where we consider a platoon of $n$ identical vehicles on a road, $s_i$ ($i = 1, ..., n$) is the back-to-back distance of the $i$-th vehicle from the $(i-1)$-th vehicle, $v_i$ ($i = 1, ..., n$) is the speed of the $i$-th vehicle and $u_i$ ($i = 1, ..., n$) is the control input (acceleration) of the $i$-th vehicle. For model (2.1), we have the following cases:

- $v_0$ is the speed of the leader and is an external input. This corresponds to the case of an open road,

- $v_0 = v_n$, which corresponds to the case of a ring-road. In this case, the identity $\sum_{i=1}^{n} s_i = L$ holds, where $L > 0$ is the length of the ring-road.

For autonomous vehicles (no communication), the so-called Predecessor-Following control architecture is used, i.e., there exists a function $F : \mathfrak{R}_+^3 \to \mathfrak{R}$ so that

$$u_i = F(s_i, v_{i-1}, v_i), \ i = 1,...,n. \tag{2.2}$$

The function $F : \mathfrak{R}_+^3 \to \mathfrak{R}$ is a feedback law that constitutes the Adaptive Cruise Controller. The function must be selected in such a way that the following requirements hold.

<u>1) Safe Operation Requirement for the open road case:</u> *There exists constant $a > 0$, a non-empty set of inputs $J \subseteq \{ v_0 \in C^1(\mathfrak{R}_+) : 0 < v_0 < v_{\max} \}$, where $v_{\max} > 0$ is the speed limit of the road, and a set valued map $(0, v_{\max}) \ni v_0 \to D(v_0) \subseteq \mathfrak{R}^{2n}$ with*

$$D(v_0) \subseteq \{ (s_1,...,s_n, v_1,...,v_n) \in \mathfrak{R}^{2n} : 0 < v_i < v_{\max}, s_i > a, i = 1,...,n \} \tag{2.3}$$

*with the following property:*

"*For each $v_0 \in J$, if $(s_1(0),...,s_n(0), v_1(0),...,v_n(0)) \in D(v_0(0))$, then the solution of the initial-value problem (2.1), (2.2) with initial condition $(s_1(0),...,s_n(0), v_1(0),...,v_n(0))$ exists for all $t \geq 0$ and satisfies $(s_1(t),...,s_n(t), v_1(t),...,v_n(t)) \in D(v_0(t))$.*"

Notice that requirement of safe operation is actually a well-posedness requirement, i.e., we require that the solution exists and takes values on a physically meaningful set. However, the requirement of safe operation is not only a well-posedness characterization of the solution; we further require that $s_i(t) > a$, where the constant $a > 0$ is the minimum allowable distance of two vehicles. This is a safety requirement which implies the absence of collisions.

For the ring-road case, the safe operation requirement takes the following form when $L > na$ (an essential constraint which guarantees that the vehicles can be placed in the ring-road).

<u>1') Safe Operation Requirement for the ring-road case:</u> *There exist constants $a > 0$, $v_{\max} > 0$ and a set $D \subseteq \mathfrak{R}^{2n}$ with*

$$D \subseteq \{ (s_1,...,s_n, v_1,...,v_n) \in \mathfrak{R}^{2n} : 0 < v_i < v_{\max}, s_i > a, i = 1,...,n \} \tag{2.4}$$

*with the following property:*

"*If $(s_1(0),...,s_n(0), v_1(0),...,v_n(0)) \in D$ and $\sum_{i=1}^{n} s_i(0) = L$, then the solution of the initial-value problem (2.1), (2.2) with $v_0 = v_n$, initial condition $(s_1(0),...,s_n(0), v_1(0),...,v_n(0))$ exists for all $t \geq 0$ and satisfies $(s_1(t),...,s_n(t), v_1(t),...,v_n(t)) \in D$.*"

2) Technical Requirement: *For a given constant $A > 0$, we have*

$$|F(s,w,v)| \leq A, \text{ for all } s > a, \ v, w \in (0, v_{\max}). \tag{2.5}$$

The constant $A > 0$ appearing in the technical requirement is the maximum acceleration that the vehicle can have and depends on the technical characteristics of the vehicles and the road.

3) Stability Requirement for the open road: *For every $v^* \in (0, v_{\max})$, there exists $s^* \in (a, +\infty)$ with $F(s^*, v^*, v^*) = 0$ such that (i) $(s^*,...,s^*,v^*,...,v^*) \in D(v^*)$, (ii) the constant input $v_0(t) \equiv v^*$ is in the allowable input set $J$, and (iii) the equilibrium point $(s^*,...,s^*,v^*,...,v^*) \in D(v^*)$ of (2.1), (2.2) with $v_0(t) \equiv v^*$ defined on $\overline{D(v^*)}$ is Globally Asymptotically Stable and Locally Exponentially Stable, i.e., there exist constants $M, \sigma, \delta > 0$ and a function $\omega \in KL$ so that for every $(s_1(0),...,s_n(0),v_1(0),...,v_n(0)) \in \overline{D(v^*)}$ the solution of (2.1), (2.2) with $v_0(t) \equiv v^*$ satisfies*

$$\left|\left(s_1(t) - s^*,...,s_n(t) - s^*, v_1(t) - v^*,...,v_n(t) - v^*\right)\right|$$
$$\leq \omega\left(\left|\left(s_1(0) - s^*,...,s_n(0) - s^*, v_1(0) - v^*,...,v_n(0) - v^*\right)\right|, t\right)$$
$$\text{for all } t \geq 0; \tag{2.6}$$

*and if in addition $\left|\left(s_1(0) - s^*,...,s_n(0) - s^*, v_1(0) - v^*,...,v_n(0) - v^*\right)\right| < \delta$ then*

$$\left|\left(s_1(t) - s^*,...,s_n(t) - s^*, v_1(t) - v^*,...,v_n(t) - v^*\right)\right|$$
$$\leq M \exp(-\sigma t) \left|\left(s_1(0) - s^*,...,s_n(0) - s^*, v_1(0) - v^*,...,v_n(0) - v^*\right)\right|$$
$$\text{for all } t \geq 0. \tag{2.7}$$

The stability requirement is a crucial requirement that guarantees the convergence of the vehicle states to the desired values. For a ring-road, the stability requirement takes the following form.

3') Stability Requirement for the ring-road: *There exists $v^* \in (0, v_{\max})$ with $F(s^*, v^*, v^*) = 0$, where $s^* = L/n$, such that (i) $(s^*,...,s^*,v^*,...,v^*) \in D$, and (ii) the equilibrium point $(s^*,...,s^*,v^*,...,v^*) \in D$ of (2.1), (2.2) with $v_0 = v_n$ is Globally Exponentially Stable, i.e., there exist constants $M, \sigma > 0$ so that for every $(s_1(0),...,s_n(0),v_1(0),...,v_n(0)) \in \overline{D}$ with $\sum_{i=1}^{n} s_i(0) = L$, the solution of (2.1), (2.2) with $v_0 = v_n$ satisfies estimate (2.7).*

Notice the difference in the stability requirements for an open road and for a ring-road. In a ring-road all states in the set $D$ are automatically bounded, while this is not true for the states in the set $\overline{D(v^*)}$.

While the stability requirement guarantees the desired asymptotic behavior, there is no guarantee for the transient behavior. A performance requirement which guarantees improved transient behavior is the requirement of string stability. Here we adopt a slightly stronger version of the $L_p$ string stability notion given in [19]. As noted in [19], the $L_p$ string stability notion is motivated by the requirement of energy dissipation along the string of vehicles for $p=2$, whereas the case $p=\infty$ is related to maximum overshoot of the local error vector between the current speed and desired speed.

4) <u>String Stability Requirement:</u> *There exists $p \in [1, +\infty]$ with the following property:*

*"For every $q > 0$ there exists a continuous function $\beta_q : \Re^2 \to \Re_+$ with $\beta_q(0) = 0$, $\beta_q(s) > 0$ for $s \in \Re^2 \setminus \{0\}$, such that every solution of (2.1), (2.2) with $v_0 \in J$ in the open road case and $v_0 = v_n$ in the ring-road case, satisfies the estimate*

$$\|v_i\|_{[0,t],p} \leq (1+q)\|v_{i-1}\|_{[0,t],p} + \beta_q\left(s_i(0) - s^*, v_i(0) - v^*\right)$$
*for all $t \geq 0$ and $i = 1, ..., n$"* (2.8)

*where* $\|v_i\|_{[0,t],p} = \left(\int_0^t |v_i(l) - v^*|^p \, dl\right)^{1/p}$, $\|v_{i-1}\|_{[0,t],p} = \left(\int_0^t |v_{i-1}(l) - v^*|^p \, dl\right)^{1/p}$ *for* $p \in [1, +\infty)$,

$\|v_i\|_{[0,t],\infty} = \sup_{0 \leq l \leq t}\left(|v_i(l) - v^*|\right)$, $\|v_{i-1}\|_{[0,t],\infty} = \sup_{0 \leq l \leq t}\left(|v_{i-1}(l) - v^*|\right)$, $v^* \in (0, v_{\max})$, $s^* \in (a, +\infty)$ *are constants with $F(s^*, v^*, v^*) = 0$ ($s^* = L/n$ in the case of ring-road).*

Another performance guarantee can be obtained by the existence of a globally exponentially stable manifold for the speed states. This requirement is described below.

5) <u>Fundamental Diagram Requirement:</u> *There exists a function $G \in C^1(\Re_+; \Re_+)$ and constants $\bar{M}, \bar{\sigma} > 0$ such that every solution of (2.1), (2.2) with $v_0 \in J$ in the open road case and $v_0 = v_n$ in the ring-road case, satisfies the estimate*

$$\sum_{i=1}^n |v_i(t) - G(s_i(t))| \leq \bar{M} \exp(-\bar{\sigma} t) \sum_{i=1}^n |v_i(0) - G(s_i(0))|, \text{ for all } t \geq 0.$$ (2.9)

The fundamental diagram requirement essentially demands that the vehicle speeds ultimately depend only on the local vehicle density. Since the vehicle density $\rho(t,x)$ is equal to $1/s_i(t)$ when $x$ is a position between the $i$-th vehicle from the $(i-1)$-th vehicle, it is reasonable to say that ultimately the local speed of vehicles of the platoon obeys the equation

$$v = G\left(\rho^{-1}\right), \text{ for } \rho \in (0, a^{-1}).$$ (2.10)

Even in the case that a globally exponentially manifold for the speed states is absent, it is reasonable to expect that all equilibrium points for (2.1), (2.2) satisfy a relation of the form $v_i = G(s_i)$ for $i = 1, ..., n$ and an appropriate function $G \in C^1(\Re_+; \Re_+)$. The inverse of this relation, i.e., the equation $s_i = G^{-1}(v_i)$ when $G$ is invertible, is called a spacing policy (see [26], [31]). A spacing policy allows the reduction of the study of the system of $n$ ODEs (2.1), (2.2) to the standard LWR model with speed given by (2.10) (although such a reduction is problematic in the absence of a fundamental diagram for the platoon). In this case, the following macroscopic stability condition arises.

<u>6) Macroscopic Stability Requirement:</u> *There exist constants $0 < a < b$ such that*

$$\frac{\partial}{\partial \rho}\left(\rho G\left(\rho^{-1}\right)\right) > 0, \text{ for all } \rho \in (a,b) \quad (2.11)$$

Inequality (2.11) was proposed in [26], [31] for the so-called "unconditional traffic-flow stability", i.e., the stability of the model to all possible boundary conditions. It was later used in [23] for the construction of macroscopically stable spacing policies.

A very common spacing policy used in ACC systems is the constant time-gap policy (CTG) in which the desired spacing $s_d$ is proportional to speed:

$$s_d = r + Tv_i \quad (2.12)$$

where $r \geq a$ is a safety or desired distance between vehicles and the constant $T > 0$ is referred to as the time-gap, i.e., the time required for the following vehicle to reach the back side of the front vehicle while driving with its current speed $v_i$. For the CTG spacing policy (2.12), a typical control law (2.2) to regulate the spacing between vehicles is given by

$$F(s,w,v) = (k-g)g(s-r) + gw - kv \quad (2.13)$$

where $k > g > 0$, the time-gap being $T = 1/g$, see [14], [20]. The CTG policy (2.12), with the controller (2.2), (2.13) satisfies both the Stability Requirement and the String-Stability Requirement, see [20]. However, the Technical Requirement is not fulfilled since $F(s,w,v)$ in (2.13) grows linearly in $s$ and, more importantly, there are cases where the Safe Operation Requirement on an open road may not be valid. To our knowledge, no researcher has ever shown what is the allowable set of inputs for an open road. This is illustrated in the following scenario.

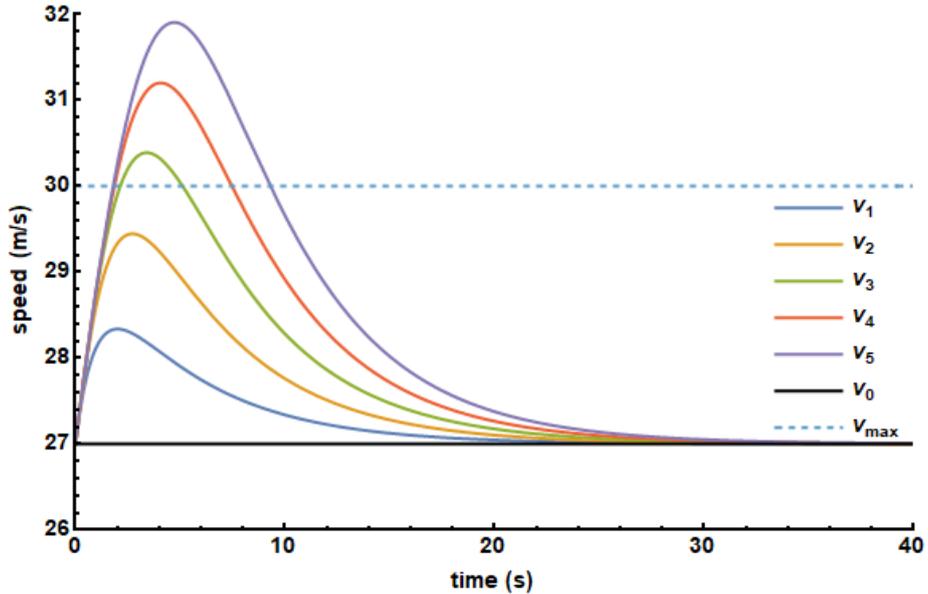

**Figure 1:** CTG policy (2.12) with controller (2.2), (2.13) and speeds exceeding the road speed limit $v_{\max} = 30.1 m/s$.

For illustration of some of the above statements, consider a case of $n = 5$ vehicles of the same length $a = 5m$ moving on a road with speed limit $v_{\max} = 30.1 m/s$ with all vehicles using the same CTG spacing policy (2.12) with controller (2.2), (2.13), initial speed $v_{i,0} = v_i(0) = 27 m/s$ and initial spacing $s_{i,0} = s_i(0) = 68m$, $i = 1,...,5$. Furthermore, suppose that the leading vehicle is also moving with constant speed $v_0 = 27 m/s$, and let the time-gap be $T = 1/g = 1s$, and $k = 1.2 s^{-1}$, $r = 31m$. Figure 1 shows that in this setting, certain vehicles do not respect the speed limit $v_{max} = 30.1 m/s$ of the road.

On a second scenario, consider the same set-up as before with initial speed $v_{i,0} = 27 m/s$, $i = 0,1,...,5$, but with initial spacing $s_{i,0} = 20m$, $i = 1,...,5$. Furthermore, suppose that the leading vehicle decelerates strongly to a significantly lower speed $v_0 = 5.4 m/s$. Figure 2 illustrates that, also in this scenario, the Safe Operation Requirement is not satisfied since certain vehicles attain negative speeds.

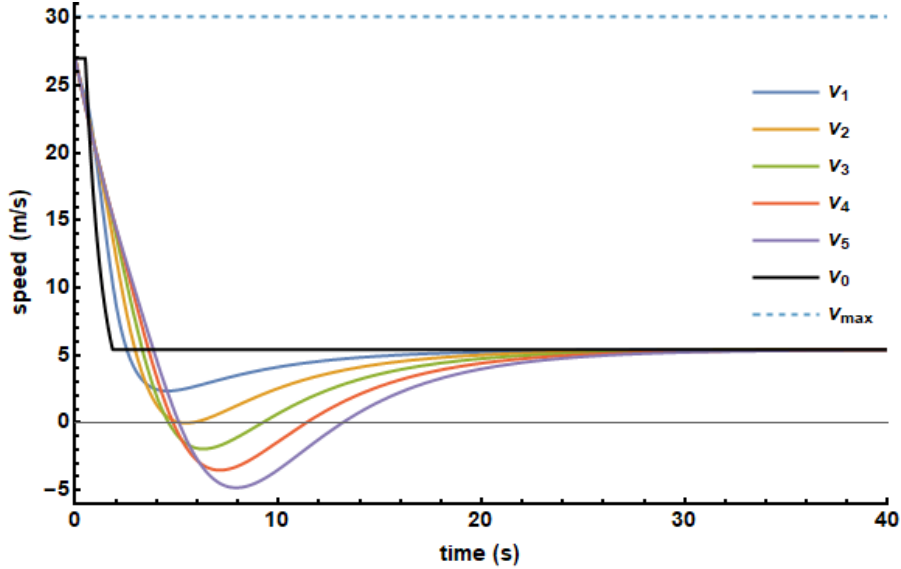

**Figure 2:** CTG policy (2.12) with controller (2.2), (2.13) with the leading vehicle strongly decelerating.

As a third scenario, we consider a slowly moving leading vehicle $v_0 = 10 m/s$ on a road with speed limit $v_{max} = 30.1 m/s$ and $n = 5$ vehicles moving with speed $v_{i,0} = 30 m/s$, $i = 1,...,5$. and initial spacing $s_{1,0} = 25m$, $s_{i,0} = 15m$, $i = 2,...,5$. For this scenario, we let the time-gap be $T = 1/g = 1s$, $k = 1.2$, as before, and set $r = 33m$; furthermore, suppose that the leading vehicle decelerates to a speed of $v_0 = 1 m/s$. Figure 3 shows the back-to-back vehicle distances for this particular scenario. It can be seen that the safe operation requirement with $a = 5m$ (the vehicles' length) is again not satisfied, since there exists time $T > 0$ with $s_2(T) < a$, which implies collision between the first and second vehicle.

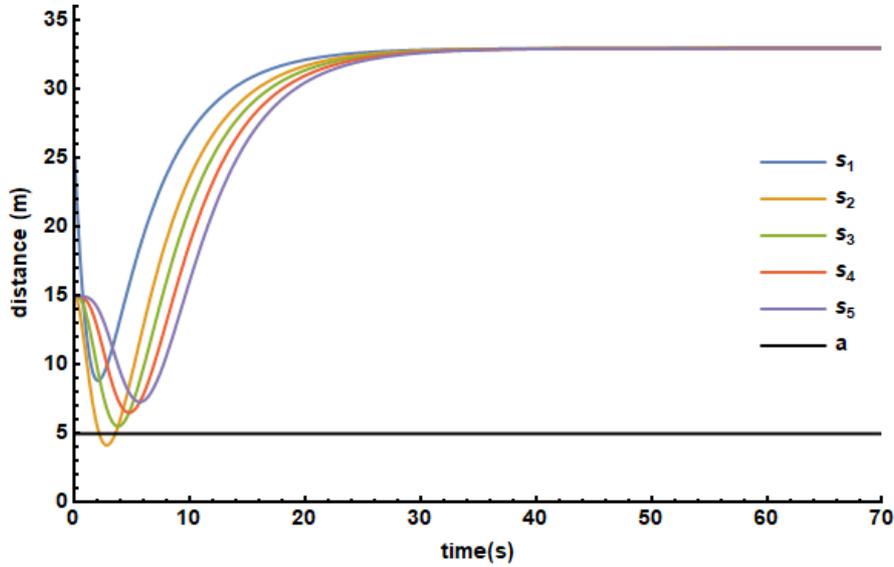

**Figure 3:** Vehicle spacing for CTG policy (2.12) with controller (2.2), (2.13) with collision.

In addition to the above scenarios, there are certain macroscopic properties of the CTG policy for a string of vehicles on a single-lane highway that are of interest. More specifically, for the CTG policy (2.12), we can obtain from (2.10) with $G(s) = g(s-r)$, that the road speed in terms of density is

$$v = g \frac{1-r\rho}{\rho} \qquad (2.14)$$

and the traffic flow is

$$Q = \rho v = g(1-r\rho). \qquad (2.15)$$

Notice now that, as the density decreases, the speed grows unbounded. Conversely, larger values of $r$ result in smaller traffic density with the speed being negative if $\rho \in (r^{-1}, a^{-1})$. Figure 4 below illustrates the density-flow relation (the so-called fundamental diagram) for different values of the minimum distance $r$ and the time-gap $T = 1/g$. It can be seen that the fundamental diagram violates the maximum velocity since it passes above the line $Q = \rho v_{max}$. It is clear that the macroscopic stability requirement does not hold (as was already remarked in [26], [31]). Moreover, since the fundamental diagram is always a straight line, the CTG policy (2.12) has limited degrees of freedom for the optimal selection of the desired fundamental diagram.

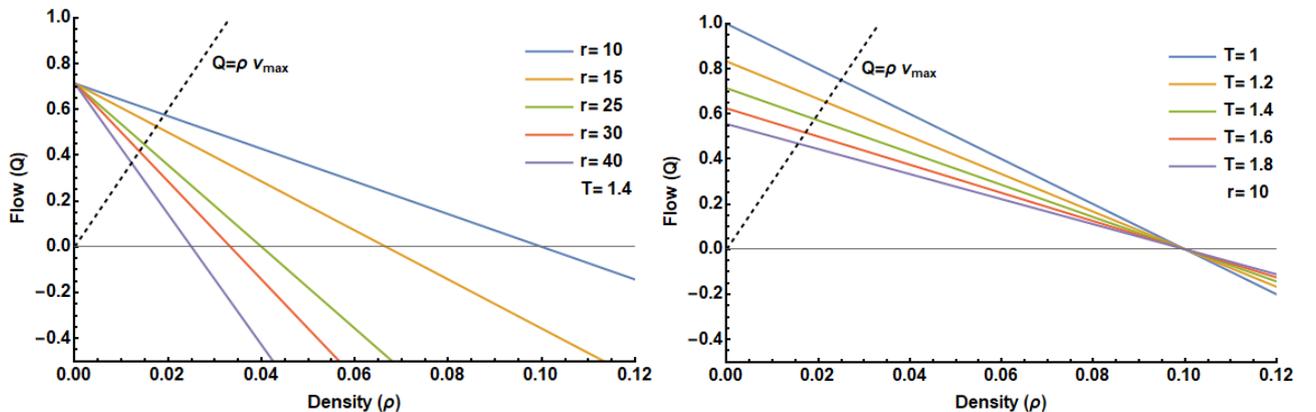

**Figure 4:** Fundamental diagram of CTG policy (2.12) for several values of $r$ and fixed $T = 1.4s$ on the left; on the right, the fundamental diagram for fixed $r = 10m$ and several values of $T$.

To summarize, we have seen three scenarios where the CTG policy (2.12) with the controller (2.13) fails to satisfy the safe requirement operation leading to negative speeds, collisions and speeds exceeding the road speed limits. The same behavior concerning safety has also been observed with the Variable Time Gap policy under the controller proposed in [31]. Due to space constraints, the corresponding simulations are not included here.

## 3. Safe Operation of Platoons

In this section, we provide sufficient conditions for the safe operation of a vehicular platoon. Due to the technical differences and structure of a platoon operating on an open road versus a ring-road, we will treat each case separately. Our first result provides sufficient conditions for an open road and is given below.

**Theorem 1 (Safe Operation in an Open Road):** *Let* $f, g, \kappa : \Re \to \Re_+$ *be locally Lipschitz functions and suppose that there exist constants* $v_{\max} > 0$, $\lambda > a > 0$, $k > 0$ *for which the functions* $f, g, \kappa : \Re \to \Re_+$ *satisfy the following properties:*

$$0 \leq g(s) < \kappa(s) \leq k, \text{ for all } s \geq a \quad (3.1)$$

$$\frac{f(s)}{\kappa(s) - g(s)} \leq v_{\max} < k(\lambda - a), \text{ for all } s \geq a \quad (3.2)$$

$$f(s) = g(s) = 0 \text{ and } \kappa(s) = k, \text{ for all } s \in [a, \lambda] \quad (3.3)$$

*Given* $v_0 \in (0, v_{\max})$, *we define the set:*

$$D(v_0) = \left\{ (s_1, \ldots, s_n, v_1, \ldots, v_n) \in \Re^{2n} : \begin{array}{c} 0 < v_i < v_{\max} \\ s_i > a + k^{-1} \max(0, v_i - v_{i-1}) \end{array}, i = 1, \ldots, n \right\}. \quad (3.4)$$

*Then for every input* $v_0 \in C^1(\Re_+)$ *satisfying*

$$\dot{v}_0(t) \geq -k v_0(t), \ 0 < v_0(t) < v_{\max}, \text{ for all } t \geq 0 \quad (3.5)$$

*and for every* $(s_{1,0}, \ldots, s_{n,0}, v_{1,0}, \ldots, v_{n,0}) \in D(v_0(0))$, *the initial-value problem (2.1), (2.2) with*

$$F(s, w, v) = f(s) + g(s)w - \kappa(s)v, \text{ for all } s, v, w \in \Re \quad (3.6)$$

*with initial condition* $(s_1(0), \ldots, s_n(0), v_1(0), \ldots, v_n(0)) = (s_{1,0}, \ldots, s_{n,0}, v_{1,0}, \ldots, v_{n,0})$ *has a unique solution* $(s_1(t), \ldots, s_n(t), v_1(t), \ldots, v_n(t))$ *defined for all* $t \geq 0$ *that satisfies* $(s_1(t), \ldots, s_n(t), v_1(t), \ldots, v_n(t)) \in D(v_0(t))$ *for all* $t \geq 0$.

Theorem 1 characterizes clearly the class of inputs that can be allowed for the safe operation of a vehicular platoon. Indeed, the speed of the leader $v_0$ must be a function of class $C^1(\Re_+)$, which satisfies (3.5). When the speed of the leader satisfies this safety requirement, then all vehicles remain in a distance at least $a > 0$ from each other, and all vehicles' speeds are less than the speed limit $v_{\max}$. Thus, if the adaptive cruise controller has the form (3.6), where the functions $f, g, \kappa : \Re \to \Re_+$ satisfy (3.1), (3.2) and (3.3), then the safe operation requirement is satisfied.

Notice that the sufficient conditions (3.1), (3.2) and (3.3), are not restrictive and depend on technical characteristics of the vehicles and the road. In particular, the constant $k$ in (3.1) represents a friction term and condition (3.5) together with inequality $\kappa(s) \leq k$, $s \geq a$, describe the maximum rate of deceleration of the leading and following vehicles in the platoon, respectively. Condition (3.3) describes the distance at which a following vehicle starts decelerating. Finally, conditions (3.1) and (3.2) are technical conditions that are required for the safe operation of the platoon.

If the adaptive cruise controller has the form (3.6), where the functions $f, g, \kappa : \Re \to \Re_+$ satisfy (3.1), (3.2) and (3.3), then the same safety requirements are guaranteed even in the case of a ring-road. This is shown by the following theorem.

**Theorem 2 (Safe Operation in a Ring-Road):** *Let $f, g, \kappa : \Re \to \Re_+$ be locally Lipschitz functions and suppose that there exist constants $v_{\max} > 0$, $\lambda > a > 0$, $k > 0$ for which the functions $f, g, \kappa : \Re \to \Re_+$ satisfy (3.1), (3.2) and (3.3). Define the set:*

$$D = \left\{ (s_1, ..., s_n, v_1, ..., v_n) \in \Re^{2n} : \begin{array}{c} 0 < v_i < v_{\max} \\ s_i > a + k^{-1} \max(0, v_i - v_{i-1}) \end{array}, i = 1, ..., n, v_0 = v_n \right\} \quad (3.7)$$

*Then for every $(s_{1,0}, ..., s_{n,0}, v_{1,0}, ..., v_{n,0}) \in D$ with $\sum_{i=1}^{n} s_{i,0} = L$, the initial-value problem (2.1), (2.2) with (3.6), $v_0 \equiv v_n$ and initial condition $(s_1(0), ..., s_n(0), v_1(0), ..., v_n(0)) = (s_{1,0}, ..., s_{n,0}, v_{1,0}, ..., v_{n,0})$ has a unique solution $(s_1(t), ..., s_n(t), v_1(t), ..., v_n(t))$ defined for all $t \geq 0$ that satisfies $(s_1(t), ..., s_n(t), v_1(t), ..., v_n(t)) \in D$ and $\sum_{i=1}^{n} s_i(t) = L$ for all $t \geq 0$.*

**Remark:** If the adaptive cruise controller has the form (3.6), where the functions $f, g, \kappa : \Re \to \Re_+$ satisfy (3.1), (3.2) and (3.3), then the Technical Requirement holds for the function $F$ defined by (3.6). Indeed, the fact that the functions $f, g, \kappa : \Re \to \Re_+$ are non-negative and inequality (3.2) show that

$$|F(s, w, v)| < k v_{\max}, \text{ for all } s > a, \ v, w \in (0, v_{\max}) \quad (3.8)$$

Consequently, inequality (3.8) guarantees that inequality (2.5) holds with $A := k v_{\max}$.

## 4. String Stability and Fundamental Diagram

If the adaptive cruise controller has the form (3.6), where the functions $f, g, \kappa : \Re \to \Re_+$ satisfy (3.1), (3.2) and (3.3), then the safe operation of a vehicular platoon is guaranteed. However, we have no guarantee for the string stability of the platoon or for the existence of a fundamental diagram. In order to achieve these objectives, we have to restrict the allowable form of the adaptive cruise controller so that conditions (3.1), (3.2), (3.3) hold automatically, and additional sufficient conditions that guarantee string stability and the existence of a fundamental diagram for the platoon hold. This is shown by the following theorem, which addresses the case of an open road.

**Theorem 3 (String Stability and Fundamental Diagram for an open road):** *Let $g: \Re \to \Re_+$ be a locally Lipschitz function and suppose that there exist constants $k > g_{max} > 0$, $\lambda > a > 0$ for which the following properties hold:*

$$0 < g(s) \leq g_{max}, \text{ for all } s > \lambda \tag{4.1}$$

$$v_{max} := \int_a^{+\infty} g(l)dl < k(\lambda - a) \tag{4.2}$$

$$g(s) = 0, \text{ for all } s \in [a, \lambda] \tag{4.3}$$

*Let $v^* \in (0, v_{max})$ be a given constant and define $s^* \in (\lambda, +\infty)$ by means of the equation*

$$v^* = G(s^*) \tag{4.4}$$

*where*

$$G(s) := \int_a^s g(l)dl, \text{ for all } s \in \Re \tag{4.5}$$

*Also define*

$$F(s, w, v) = (k - g(s))G(s) + g(s)w - kv, \text{ for all } s, v, w \in \Re. \tag{4.6}$$

*Given $v_0 \in (0, v_{max})$, we define the set $D(v_0) \subset \Re^{2n}$ by means of (3.4). Then for every input $v_0 \in C^1(\Re_+)$ satisfying (3.5) and for every $(s_{1,0},...,s_{n,0}, v_{1,0},...,v_{n,0}) \in D(v_0(0))$, the initial-value problem (2.1), (2.2) with (4.6), initial condition $(s_1(0),...,s_n(0), v_1(0),...,v_n(0)) = (s_{1,0},...,s_{n,0}, v_{1,0},...,v_{n,0})$ has a unique solution $(s_1(t),...,s_n(t), v_1(t),...,v_n(t))$ defined for all $t \geq 0$ that satisfies $(s_1(t),...,s_n(t), v_1(t),...,v_n(t)) \in D(v_0(t))$ for all $t \geq 0$. Moreover, the following inequalities hold for all $t \geq 0$, $i = 1,...,n$ and $q > 0$:*

$$\int_0^t \left(v_i(\tau) - v^*\right)^2 d\tau \leq (1+q) \int_0^t \left(v_{i-1}(\tau) - v^*\right)^2 d\tau$$
$$+ k^{-1} \left(W(s_i(0), v_i(0)) + \frac{1}{2q}(v_i(0) - G(s_i(0)))^2\right) \tag{4.7}$$

$$\int_0^t \left(G(s_i(\tau)) - v^*\right)^2 d\tau \leq (1+2q)\frac{2qk + k - g_{max}}{k - g_{max}} \int_0^t \left(v_{i-1}(\tau) - v^*\right)^2 d\tau$$
$$+ \frac{2qk + k - g_{max}}{k(k - g_{max})} \left(W(s_i(0), v_i(0)) + \frac{1}{2q}(v_i(0) - G(s_i(0)))^2\right) \tag{4.8}$$

$$\left|v_i(t) - v^*\right| \leq 2\left|v_i(0) - v^*\right| + \left|G(s_i(0)) - v^*\right| + \sup_{0 \leq \tau \leq t}\left(\left|v_{i-1}(\tau) - v^*\right|\right) \tag{4.9}$$

$$\sum_{i=1}^n \left|v_i(t) - G(s_i(t))\right| \leq \exp(-(k - g_{max})t) \sum_{i=1}^n \left|v_i(0) - G(s_i(0))\right| \tag{4.10}$$

*where $W(s, v) := (v - v^*)^2 + 2\int_{s^*}^s (k - g(z))(G(z) - v^*)dz$.*

Due to (4.1), (4.2), (4.3), the function $G$, defined by (4.5), is strictly increasing on $[\lambda,+\infty)$. This feature guarantees that for every $v^* \in (0, v_{\max})$, the solution $s^* > \lambda$ of equation (4.4) is unique.

It should be noted that if the adaptive cruise controller has the form (4.6), where $g: \Re \to \Re_+$ is a locally Lipschitz function that satisfies (4.1), (4.2), (4.3), then the conditions for the safe operation of the vehicular platoon hold. However, in this case we also have some additional properties shown by estimates (4.7), (4.8), (4.9) and (4.10). Estimate (4.7) shows that the $L_2$ string stability notion holds; and estimate (4.9) shows that the $L_\infty$ string stability notion holds. The point $(s_1,...,s_n,v_1,...,v_n) = (s^*,...,s^*,v^*,...,v^*)$ is the desired equilibrium point for the vehicular platoon. Moreover, estimate (4.10) guarantees that the vehicular platoon under the adaptive cruise controller (4.6) has a fundamental diagram of the form (2.10), where $G$ is defined by (4.5).

Theorem 3 allows the selection of the locally Lipschitz function $g: \Re \to \Re_+$ that satisfies (4.1), (4.2), (4.3), in order to have an appropriate fundamental diagram for the platoon. By changing $g: \Re \to \Re_+$ we are in a position to change the shape as well as the critical density and the capacity of the fundamental diagram. This feature is illustrated in Section 6.

The following theorem guarantees that the same performance requirements with the open road case also hold for the case of a ring-road, when the adaptive cruise controller has the form (4.6), where $g: \Re \to \Re_+$ is a locally Lipschitz function that satisfies (4.1), (4.2), (4.3).

**Theorem 4 (String Stability and Fundamental Diagram for a Ring-Road):** *Let $g: \Re \to \Re_+$ be a locally Lipschitz function and suppose that there exist constants $k > g_{\max} > 0$, $\lambda > a > 0$ for which (4.1), (4.2), (4.3) hold. Let $v^* \in (0, v_{\max})$ be a given constant and define $s^* \in (0,+\infty)$ by means of (4.4). Define the set $D \subset \Re^{2n}$ by means of (3.7). Then for every $(s_{1,0},...,s_{n,0},v_{1,0},...,v_{n,0}) \in D$ with $\sum_{i=1}^{n} s_{i,0} = L$, the initial-value problem (2.1), (2.2) with (4.6), $v_0 \equiv v_n$, initial condition $(s_1(0),...,s_n(0),v_1(0),...,v_n(0)) = (s_{1,0},...,s_{n,0},v_{1,0},...,v_{n,0})$ has a unique solution $(s_1(t),...,s_n(t),v_1(t),...,v_n(t))$ defined for all $t \geq 0$ that satisfies $(s_1(t),...,s_n(t),v_1(t),...,v_n(t)) \in D$ and $\sum_{i=1}^{n} s_i(t) = L$ for all $t \geq 0$. Moreover, inequalities (4.7), (4.8), (4.9) and (4.10) hold for all $t \geq 0$, $i = 1,...,n$ and $q > 0$, where $W(s,v) := (v-v^*)^2 + 2\int_{s^*}^{s} (k-g(z))(G(z)-v^*)dz$.*

## 5. Stability

If the adaptive cruise controller has the form (4.6), where $g: \Re \to \Re_+$ is a locally Lipschitz function that satisfies (4.1), (4.2), (4.3), then the equilibrium point $(s^*,...,s^*,v^*,...,v^*) \in D(v^*)$ for a platoon on an open road is Globally Asymptotically Stable. In other words, the sufficient conditions for string stability and the existence of a fundamental diagram also guarantee global asymptotic stability of the equilibrium point. This is guaranteed by the following theorem.

**Theorem 5 (Stability for Open Road):** *Let $g: \Re \to \Re_+$ be a locally Lipschitz function for which there exist constants $k > g_{\max} > 0$, $\lambda > a > 0$ such that properties (4.1), (4.2), (4.3) hold. Consider a platoon of $n$ vehicles on a open/straight road described by (2.1), (2.2) with (4.6),* $v_0 = v^* \in (0, v_{\max})$ *being the constant speed of the leading vehicle, defined on the set* $\overline{D(v^*)}$, *where* $D(v^*)$ *is given by (3.4) with* $v_0 = v^* \in (0, v_{\max})$. *Define also* $s^* \in (\lambda, +\infty)$ *by means of equation (4.4). Then, the equilibrium point* $(s^*, ..., s^*, v^*, ..., v^*)$ *is Globally Asymptotically Stable. Moreover, if in addition* $g$ *is of class* $C^1$ *in a neighborhood of* $s^* > \lambda$, *then the equilibrium point* $(s^*, ..., s^*, v^*, ..., v^*)$ *is Locally Exponentially Stable.*

Theorem 5 shows that the only additional requirement for local exponential stability is a mild regularity assumption; namely, that $g$ has to be of class $C^1$ in a neighborhood of $s^* > \lambda$. However, when we study the vehicular platoon in a ring-road, then additional assumptions have to hold. Define for $n = 2, 3, ...$

$$\mu_n = \min \left\{ \sum_{i=1}^{n} (x_i - x_{i-1})^2 : x = (x_1, ..., x_n) \in \Re^n, x_0 = x_n, |x| = 1, \sum_{i=1}^{n} x_i = 0 \right\} \quad (5.1)$$

and notice that $\mu_n > 0$ for all $n = 2, 3, ...$ and that for every $x = (x_1, ..., x_n) \in \Re^n$ with $\sum_{i=1}^{n} x_i = 0$, it holds that

$$(x_1 - x_n)^2 + \sum_{i=2}^{n} (x_i - x_{i-1})^2 \geq \mu_n |x|^2. \quad (5.2)$$

The following theorem provides sufficient conditions for global exponential stability of the equilibrium point $(s^*, ..., s^*, v^*, ..., v^*) \in D$ for a vehicular platoon in a ring-road.

**Theorem 6 (Stability for Ring-Road):** *Let* $g : \Re \to \Re_+$ *be a locally Lipschitz function for which there exist constants* $k > g_{\max} > 0$, $\lambda > a > 0$ *such that properties (4.1), (4.2), (4.3) hold and consider* $n$ *vehicles along a ring-road of length* $L > n\lambda$ *described by the model (2.1), (2.2), (4.6) with* $v_0 = v_n$ *defined on the set*

$$\Omega = \overline{D} \cap \left\{ (s_1, ..., s_n, v_1, ..., v_n) \in \Re^{2n} : \sum_{i=1}^{n} s_i = L \right\} \quad (5.3)$$

*where the set* $D$ *is given by (3.7). Assume that there exist constants* $p > 0$, $M \in (0, p\mu_n / 4)$ *such that*

$$|G(s) - v^* - p(s - s^*)| \leq M |s - s^*|, \text{ for all } s \in [a, L - (n-1)a] \quad (5.4)$$

*where* $G$ *is defined by (4.5). Then the equilibrium point* $(s^*, ..., s^*, v^*, ..., v^*)$ *with* $s^* = L/n$, $v^* = G(s^*)$ *is Globally Exponentially Stable for system (2.1), (2.2), (4.6) with* $v_0 = v_n$ *defined on* $\Omega$.

It is clear that for global exponential stability in a ring-road we need the additional assumption (5.4) for the adaptive cruise controller. However, it should be noted that if the cruise controller has the form (4.6), where $g : \Re \to \Re_+$ is a locally Lipschitz function that satisfies (4.1), (4.2), (4.3) and (5.4), then the adaptive cruise controller satisfies all stability, performance, safety and technical requirements *both in an open road and in a ring-road*.

# 6. Illustrative Examples

In the simulation results below, we compare the three scenarios of the CTG policy presented in Section 2 with the proposed controller (2.2) with (4.6) and the function $g$ defined by

$$g(s) = \begin{cases} 0 & s \leq \lambda \\ (s-\lambda) & \lambda < s \leq g_{max} + \lambda \\ g_{max} & g_{max} + \lambda < s \leq \gamma \\ g_{max}\exp(\gamma - s) & s > \gamma \end{cases} \quad (6.1)$$

with $\gamma, \lambda > 0$ and $k > g_{max} > 0$. From (6.1), (2.10), (4.5), and (4.2), we obtain the fundamental diagram shown in Figure 5 for fixed values $\lambda = 30.5m$, $k = 1.2s^{-1}$ and different values of $\gamma$, $g_{max}$, all of which satisfy $v_{max} = 30.1m/s$ (recall (4.2)). Figure 5 illustrates the macroscopic stability requirement and the freedom of controlling the capacity flow and the critical density via corresponding ACC settings. It should be noticed that $g(\cdot)$ in (6.1) was selected for its simplicity, and can in general be selected such that the emerging fundamental diagram may be any desired curve which satisfies necessary physical and technical requirements (for example it should satisfy $Q \leq v_{max}\rho$).

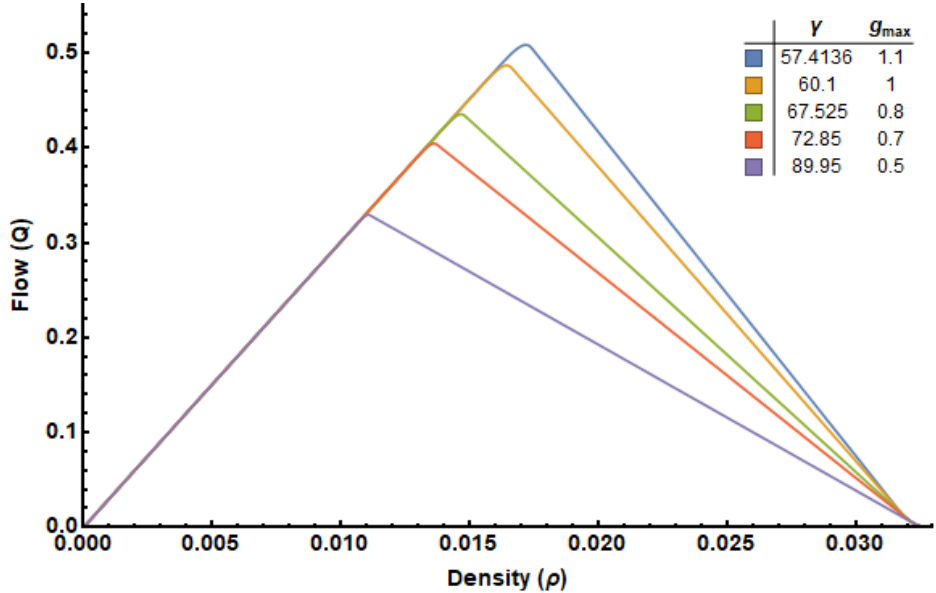

**Figure 5:** Fundamental Diagram for the nonlinear adaptive cruise controller (2.2), with (4.6) and (6.1).

For the following simulation scenarios, we consider the function $g(\cdot)$ in (6.1) with $\gamma = 62.1m$ and $g_{max} = 1s^{-1}$. For this selection, all conditions (4.1), (4.2), and (4.3) are fulfilled and, in addition, both the CTG policy (2.12) with (2.2), (2.13) and the nonlinear controller (2.2), (4.6) with (6.1) have the same speed $v^*$ and spacing equilibrium $s^*$.

**Scenario 1.** Recall that in this scenario the leading vehicle is moving with constant speed $v_0 = 27m/s$, $v_{max} = 30.1m/s$ and $v_{i,0} = 27m/s$ for $i = 0,1,...,5$ and $s_{i,0} = 68m$, $i = 1,...,5$. Notice that these initial conditions belong to the set $D(v_0)$ defined by (3.4) with $a = 5m$ for the Safe Operation requirement. Figure 6 shows the speeds of all vehicles using the adaptive cruise controller (2.2), (4.6) with (6.1). Contrary to the CTG policy (2.12) with (2.2), (2.13) (see Figure 1), the speeds of all vehicle with the nonlinear controller stay within the bounds $(0, v_{max})$. Figure 7 illustrates the vehicle spacing of the adaptive cruise controller (2.2), (4.6) with (6.1). Both Figure 6 and Figure 7 exhibit exponential convergence of the state to the equilibrium point.

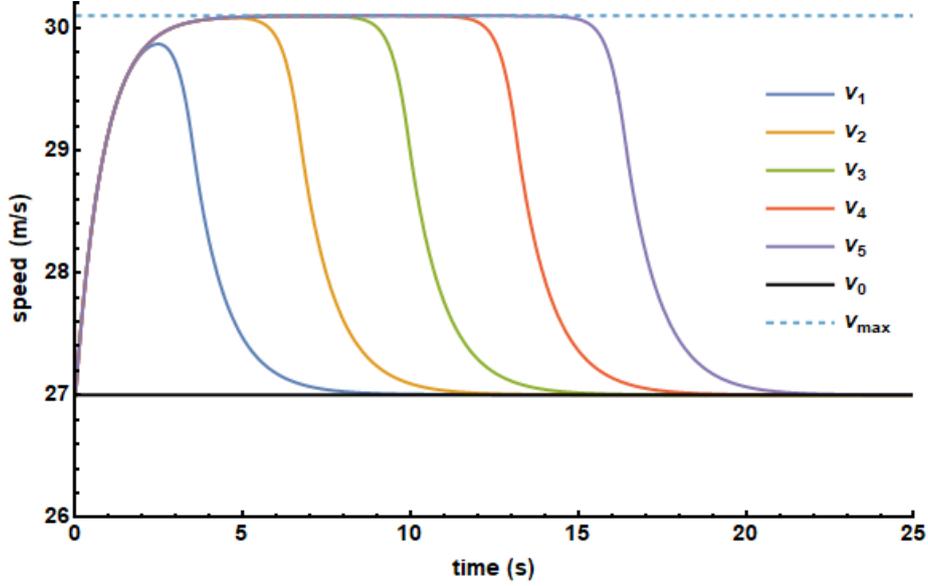

**Figure 6:** The speed of all vehicles for the nonlinear adaptive cruise controller (2.2), (4.6) with (6.1) remain within the road speed limit range.

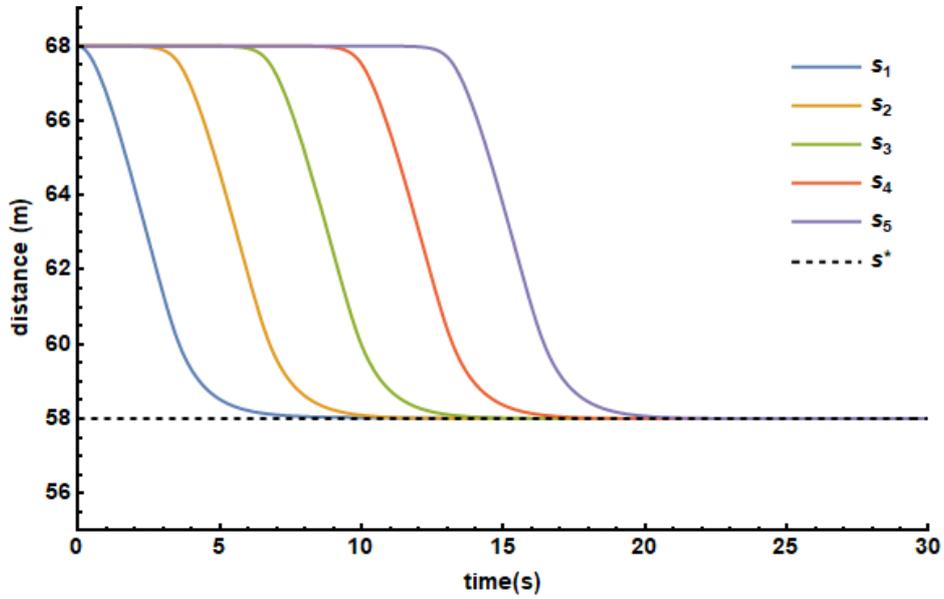

**Figure 7:** Vehicle spacing for the nonlinear adaptive cruise controller (2.2), (4.6) with (6.1).

**Scenario 2:** We focus now on the second scenario where all vehicles have initially the same speed $v_{i,0} = 27 m/s$ and the leading vehicle decelerates from the initial speed $v_0(0) = 27 m/s$ to a speed of $5.4 m/s$ with deceleration satisfying (3.8). Recall that the initial vehicle distances for this scenario are $s_{i,0} = 20m$, $i = 1,...,5$, which guarantees that the initial state is in the set $D(v_0(0))$ defined by (3.4) with $a = 5m$. The vehicle distances are shown in Figure 8, where all spacings converge exponentially to their equilibrium values.

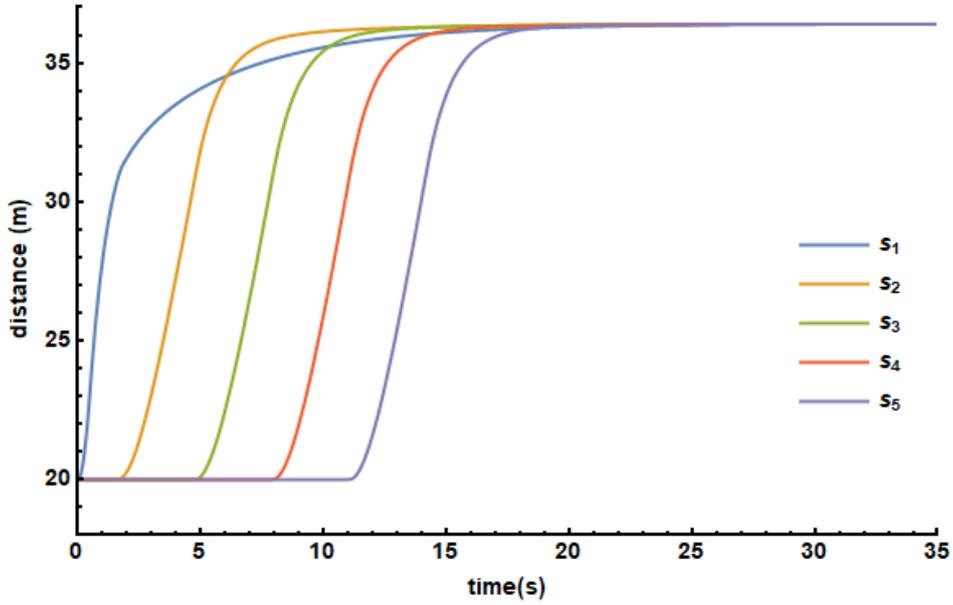

**Figure 8:** Vehicle spacing of the nonlinear adaptive cruise controller (2.2), (4.6) with (6.1).

The speed of all vehicles can be seen in Figure 9. The vehicles decelerate and retain a very slow speed satisfying $v_i(t) \in (0, v_{max})$, $i = 1,...,5$; and start accelerating to the desired speed when the distance to the preceding vehicle increases. On the contrary, using the CTG policy with the same initial conditions, the speed of the vehicles can become negative (compare with Figure 2).

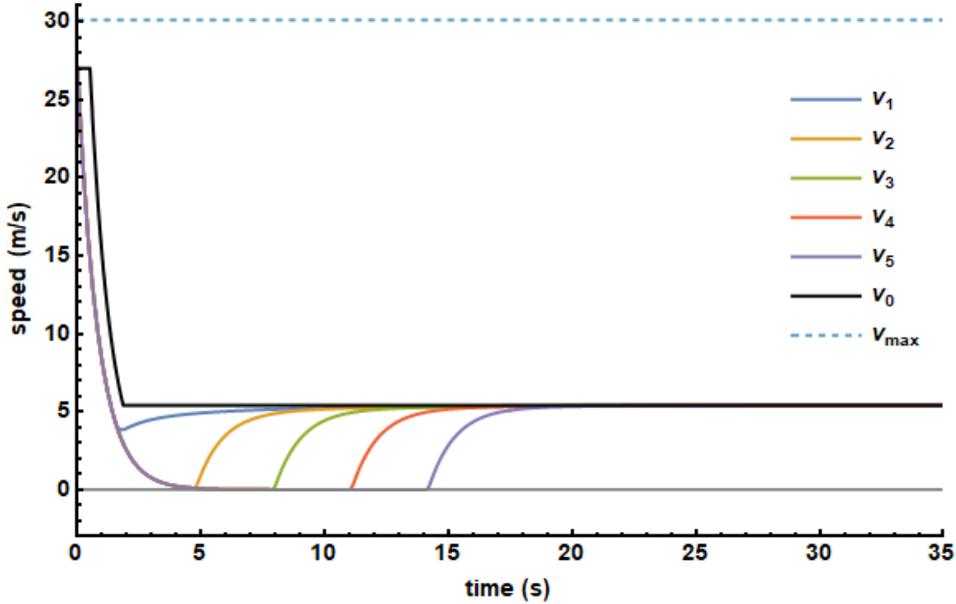

**Figure 9:** Speed of vehicles for the nonlinear adaptive cruise controller (2.2), (4.6) with (6.1), following a leader with strong deceleration.

**Scenario 3:** In this scenario, the leading vehicle has initial speed $v_0(0) = 10 m/s$ on a road with $v_{max} = 30.1 m/s$ and decelerates to $1m/s$. Recall that the initial speed and initial spacing of the $n = 5$ vehicles are $v_{i,0} = 30 m/s$, $i = 1,...,5$ and $s_{1,0} = 25m$, $s_{i,0} = 15m$, $i = 2,...,5$, respectively. Notice now that these initial conditions are in the safe operation set $D(v_0(0))$ given by (3.4). Indeed, $s_{1,0}(0) = 25 > a + k^{-1} \max(v_{1,0}(0) - v_0(0)) = 23.18m$ and $s_{i,0}(0) > 5m$ for $i = 2,...,5$. Under these initial conditions the Safe Operation requirement was not satisfied for the CTG policy (2.12) with cruise controller (2.2), (2.13), as was shown in Figure 3. On the other hand, using the proposed nonlinear adaptive cruise controller (2.2), (4.6) with (6.1), there are no collisions as shown in Figure 10. Finally, Figure 11 shows that the speeds of all vehicles remain below the speed limits, verifying the Safe Operation requirement and exponential convergence to the equilibrium point.

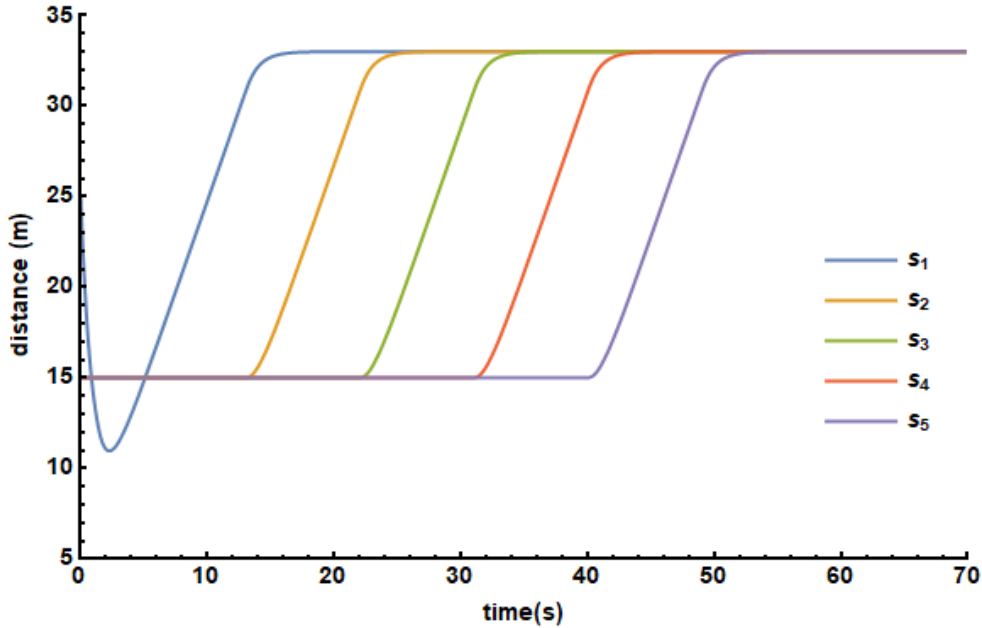

**Figure 10:** Vehicle spacing for scenario 3 using the nonlinear adaptive cruise controller (2.2), (4.6) with (6.1).

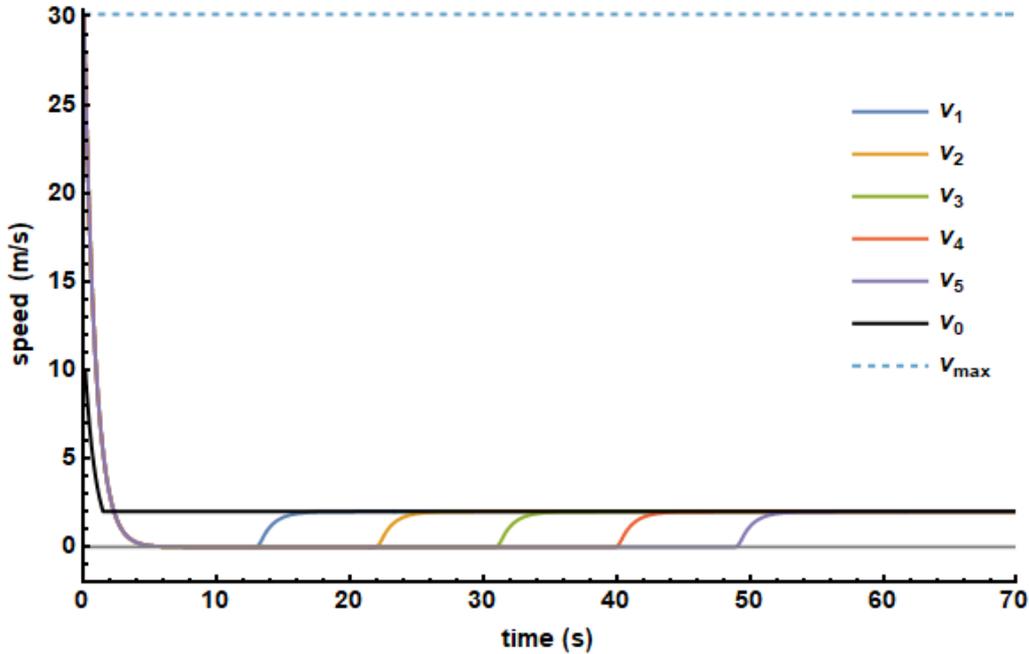

**Figure 11:** Speed of vehicles for the nonlinear adaptive cruise controller (2.2), (4.6) with (6.1) for scenario 3.

**Ring-Road Scenario:** Hereafter we consider a scenario of $n=4$ vehicles moving on a ring-road. The nonlinear controller in this case is given by (2.2), with (4.6) and (6.1) with $a=5m$, $k=2\ s^{-1}$, $\lambda=7.1m$, $\gamma=19m$, $g_{max}=0.26\ s^{-1}$, and the road length equals $L=43m$. In this scenario, we obtain from (4.2) that $v_{max}=3.32 m/s$ and the equilibrium point is $v^*=0.915 m/s$, $s^*=10.75m$.

Furthermore, we get from (5.1) that $\mu_4 = 2$, and, by setting $p = g_{max}$ and $M = 0.96 g_{max}/2$, we also guarantee that condition (5.4) is fulfilled. The simulation results are depicted in Figure 12 and Figure 13 which show the convergence to the spacing and speed equilibrium, respectively. The initial conditions for this scenario are $s_{1,0} = 10m$, $s_{2,0} = 11m$, $s_{3,0} = 12m$, $s_{4,0} = 10m$ and $v_{1,0} = 0.8m/s$, $v_{2,0} = 1.5m/s$, $v_{3,0} = 1.25m/s$, $v_{4,0} = 0.75m/s$.

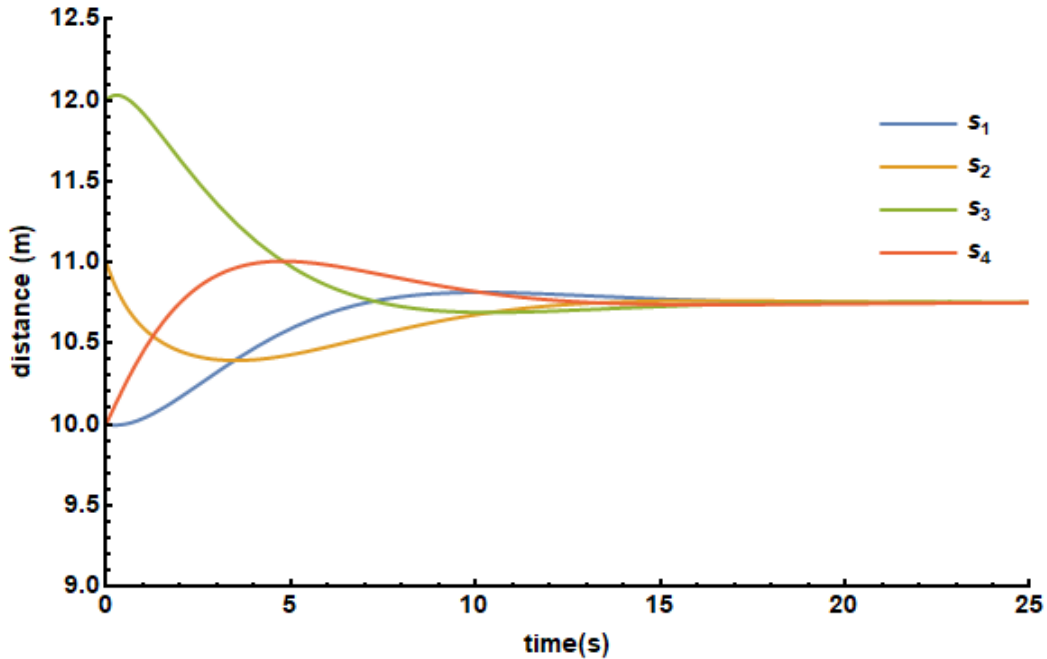

**Figure 12:** Vehicle distances on a ring-road for cruise controller (2.2), (4.6) with (6.1).

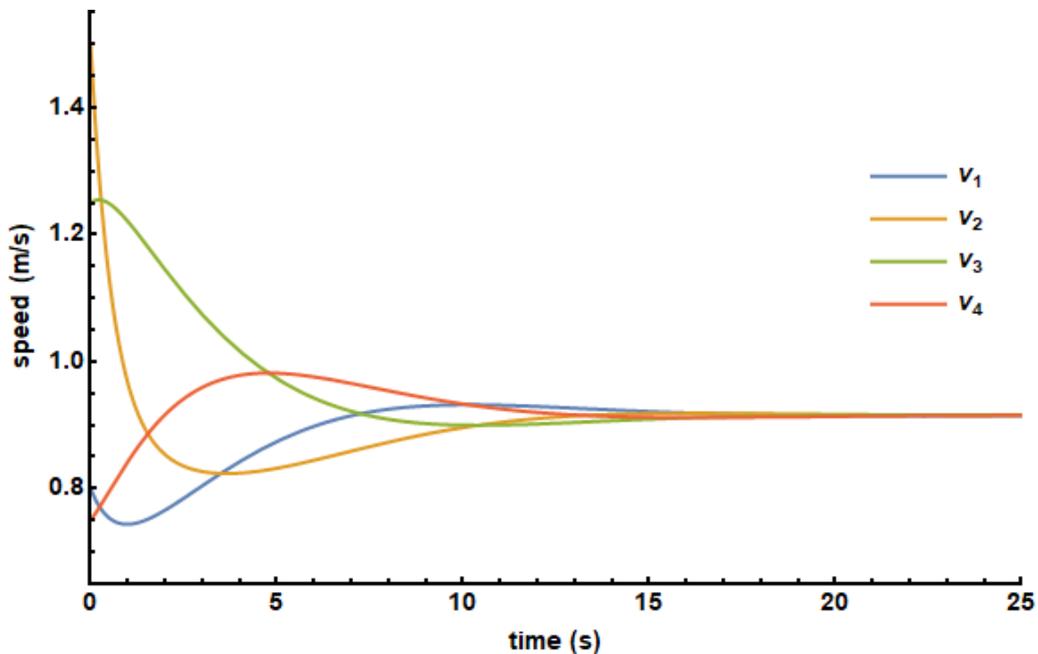

**Figure 13:** Speed of vehicles on a ring-road for cruise controller (2.2), (4.6) with (6.1).

Finally, Figure 14 depicts the evolution of the Euclidean norm of the deviation from the equilibrium $\left|\left(s_1(0) - s^*, ..., s_4(0) - s^*, v_1(0) - v^*, ..., v_4(0) - v^*\right)\right|$.

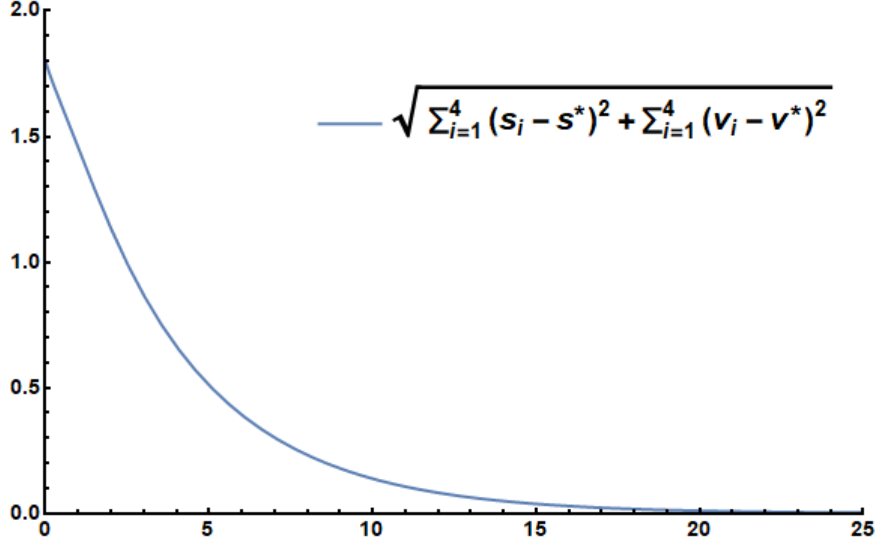

**Figure 14:** Evolution of $\left|\left(s_1(0)-s^*,...,s_4(0)-s^*,v_1(0)-v^*,...,v_4(0)-v^*\right)\right|$ for the nonlinear adaptive cruise controller (2.2), (4.6) with (6.1).

## 7. Proofs of Main Results

The proofs of Theorem 1 and Theorem 2 are similar and are performed by using a barrier function (see also [5], [36] for the use of barrier functions in control theory).

**Proof of Theorem 1:** Let an (arbitrary) input $v_0 \in C^1(\Re_+)$ that satisfies (3.5) and an (arbitrary) point $(s_{1,0},...,s_{n,0},v_{1,0},...,v_{n,0}) \in D(v_0(0))$ be given. Due to the fact that $f, g, \kappa : \Re \to \Re_+$ are locally Lipschitz functions, the initial-value problem (2.1), (2.2) with (3.6) and initial condition $(s_1(0),...,s_n(0),v_1(0),...,v_n(0)) = (s_{1,0},...,s_{n,0},v_{1,0},...,v_{n,0})$ has a unique solution $(s_1(t),...,s_n(t),v_1(t),...,v_n(t))$ defined for $t \in [0, t_{\max})$, where $t_{\max} \in (0, +\infty]$ the maximal existence time of the solution. Moreover, if $t_{\max} < +\infty$ then $\limsup_{t \to t_{\max}^-} \left(\left|(s_1(t),...,s_n(t),v_1(t),...,v_n(t))\right|\right) = +\infty$.

If $(s_1(t),...,s_n(t),v_1(t),...,v_n(t)) \in D(v_0(t))$ for all $t \in [0, t_{\max})$ then there is nothing to be proved, since (2.1), (2.2), (3.6) and (3.4) imply the differential inequalities $\dot{s}_i(t) \leq v_{\max}$ for all $t \in [0, t_{\max})$, $i = 1,...,n$. Therefore, we obtain $s_i(t) \leq s_i(0) + tv_{\max}$ for all $t \in [0, t_{\max})$, $i = 1,...,n$, and since $(s_1(t),...,s_n(t),v_1(t),...,v_n(t)) \in D(v_0(t))$, we conclude (by virtue of (3.4)) that $(s_1(t),...,s_n(t),v_1(t),...,v_n(t))$ is bounded on $[0, t_{\max})$. This implies $t_{\max} = +\infty$.

We show next by contradiction that $(s_1(t),...,s_n(t),v_1(t),...,v_n(t)) \in D(v_0(t))$ for all $t \in [0, t_{\max})$. Therefore, we next assume that there exists $t \in [0, t_{\max})$ such that $(s_1(t),...,s_n(t),v_1(t),...,v_n(t)) \notin D(v_0(t))$. By virtue of continuity of $v_0 : \Re_+ \to (0, v_{\max})$ and the fact that $(s_1(0),...,s_n(0),v_1(0),...,v_n(0)) \in D(v_0(0))$, there exists a neighborhood of $0$ such that $(s_1(t),...,s_n(t),v_1(t),...,v_n(t)) \in D(v_0(t))$ for all $t \in [0, t_{\max})$ in this neighborhood (recall (3.4)). Consequently, if we define

$$T := \inf \left\{ t \in [0, t_{\max}) : (s_1(t),...,s_n(t),v_1(t),...,v_n(t)) \notin D(v_0(t)) \right\} \qquad (7.1)$$

it follows that $T \in (0, t_{\max})$. Notice that the case $(s_1(T),...,s_n(T),v_1(T),...,v_n(T)) \in D(v_0(T))$ is excluded (since there would be a neighborhood of $T$ with $(s_1(t),...,s_n(t),v_1(t),...,v_n(t)) \in D(v_0(t))$ for all $t \in [0, t_{\max})$ in this neighborhood and that contradicts definition (7.1)). Notice also that definition (7.1) implies that

$$(s_1(t),...,s_n(t),v_1(t),...,v_n(t)) \in D(v_0(t)) \text{ for all } t \in [0, T) \tag{7.2}$$

and since $(s_1(T),...,s_n(T),v_1(T),...,v_n(T)) \notin D(v_0(T))$, we conclude from (3.4) that

$$\prod_{i=1}^{n}\left(v_i(T)(v_{\max} - v_i(T))(s_i(T) - a)(s_i(T) - a - k^{-1}(v_i(T) - v_{i-1}(T)))\right) = 0 \tag{7.3}$$

Next, define the function $\Phi : [0, T) \to \Re$ by means of the formula:

$$\Phi(t) = \sum_{i=1}^{n}\left(\frac{1}{s_i(t) - a} + \frac{1}{s_i(t) - a - k^{-1}(v_i(t) - v_{i-1}(t))} + \frac{1}{v_i(t)} + \frac{1}{v_{\max} - v_i(t)}\right) \tag{7.4}$$

Notice that (7.3) and definition (7.4) imply that $\lim_{t \to T^-}(\Phi(t)) = +\infty$. Using (2.1), (2.2), (3.6) and definition (7.4) we obtain for $t \in [0, T)$:

$$\dot{\Phi}(t) = \sum_{i=1}^{n}\left(\frac{v_i(t) - v_{i-1}(t)}{(s_i(t) - a)^2} + \frac{v_i(t) - v_{i-1}(t) + k^{-1}(\dot{v}_i(t) - \dot{v}_{i-1}(t))}{\left(s_i(t) - a - k^{-1}(v_i(t) - v_{i-1}(t))\right)^2} - \frac{\dot{v}_i(t)}{v_i^2(t)} + \frac{\dot{v}_i(t)}{(v_{\max} - v_i(t))^2}\right) \tag{7.5}$$

Using (2.1), (2.2), (3.6), (3.4), (3.1), (3.2), (3.5), and the fact that $f, g, \kappa : \Re \to \Re_+$ are non-negative functions, we obtain the following inequalities:

$$-kv_i(t) \leq \dot{v}_i(t) \leq k(v_{\max} - v_i(t)), \text{ for all } t \in [0, T), i = 1,...,n \tag{7.6}$$

Moreover, (7.2) and definition (3.4) imply that:

$$v_i(t) - v_{i-1}(t) < k(s_i(t) - a), \text{ for all } t \in [0, T), i = 1,...,n \tag{7.7}$$

By virtue of (3.2) there exists $r > 0$ so that $v_{\max} = k(\lambda - a - r)$. Using (3.5), (7.6), we get for all $t \in [0, T)$, $i = 1,...,n$:

$$v_i(t) - v_{i-1}(t) + k^{-1}(\dot{v}_i(t) - \dot{v}_{i-1}(t)) \leq v_{\max} \tag{7.8}$$

Consequently, if $s_i(t) - a - k^{-1}(v_i(t) - v_{i-1}(t)) \geq r$ for certain $i = 1,...,n$ then we get from (7.8) the inequality $\dfrac{v_i(t) - v_{i-1}(t) + k^{-1}(\dot{v}_i(t) - \dot{v}_{i-1}(t))}{\left(s_i(t) - a - k^{-1}(v_i(t) - v_{i-1}(t))\right)^2} \leq \dfrac{v_{\max}}{r^2}$. On the other hand, if $s_i(t) - a - k^{-1}(v_i(t) - v_{i-1}(t)) < r$ then $s_i(t) < a + r + k^{-1}(v_i(t) - v_{i-1}(t))$, which combined with (7.2), (3.4), (3.5) and the fact that $v_{\max} = k(\lambda - a - r)$ gives $s(t) < a + r + k^{-1}v_{\max} = \lambda$. Therefore, in this case we get from (7.6) and (2.1), (2.2), (3.6), (3.3), (3.5):

$$v_i(t) - v_{i-1}(t) + k^{-1}\left(\dot{v}_i(t) - \dot{v}_{i-1}(t)\right) \leq v_i(t) + k^{-1}\dot{v}_i(t)$$
$$= k^{-1}f(s_i(t)) + k^{-1}g(s_i(t))v_{i-1}(t) = 0$$

Consequently, in any case, we obtain for all $t \in [0, T)$, $i = 1, ..., n$:

$$\frac{v_i(t) - v_{i-1}(t) + k^{-1}\left(\dot{v}_i(t) - \dot{v}_{i-1}(t)\right)}{\left(s_i(t) - a - k^{-1}(v_i(t) - v_{i-1}(t))\right)^2} \leq \frac{v_{\max}}{r^2} \qquad (7.9)$$

Therefore, we obtain from (7.4), (7.5), (7.6), (7.7) and (7.9) for all $t \in [0, T)$:

$$\dot{\Phi}(t) \leq k \sum_{i=1}^{n}\left(\frac{1}{s_i(t) - a} + \frac{1}{v_i(t)} + \frac{1}{v_{\max} - v_i(t)}\right) + n\frac{v_{\max}}{r^2} \leq k\Phi(t) + n\frac{v_{\max}}{r^2} \qquad (7.10)$$

The differential inequality (7.10) implies that

$$\Phi(t) \leq \exp(kt)\Phi(0) + n\frac{v_{\max}}{kr^2}\left(\exp(kt) - 1\right), \text{ for all } t \in [0, T) \qquad (7.11)$$

Estimate (7.11) contradicts the implication $\lim_{t \to T^-}(\Phi(t)) = +\infty$. The proof is complete. ◁

**Proof of Theorem 2:** Let an (arbitrary) point $(s_{1,0}, ..., s_{n,0}, v_{1,0}, ..., v_{n,0}) \in D$ with $\sum_{i=1}^{n} s_{i,0} = L$ be given. Due to the fact that $f, g, \kappa : \Re \to \Re_+$ are locally Lipschitz functions, the initial-value problem (2.1), (2.2) with (3.6), $v_0 \equiv v_n$ and initial condition $(s_1(0), ..., s_n(0), v_1(0), ..., v_n(0)) = (s_{1,0}, ..., s_{n,0}, v_{1,0}, ..., v_{n,0})$ has a unique solution $(s_1(t), ..., s_n(t), v_1(t), ..., v_n(t))$ defined for $t \in [0, t_{\max})$, where $t_{\max} \in (0, +\infty]$ the maximal existence time of the solution. Moreover, if $t_{\max} < +\infty$ then $\limsup_{t \to t_{\max}^-}\left(|(s_1(t), ..., s_n(t), v_1(t), ..., v_n(t))|\right) = +\infty$. The solution also satisfies $\sum_{i=1}^{n} s_i(t) = L$ for $t \in [0, t_{\max})$.

If $(s_1(t), ..., s_n(t), v_1(t), ..., v_n(t)) \in D$ for all $t \in [0, t_{\max})$ then there is nothing to be proved, since (3.7) and the fact that $\sum_{i=1}^{n} s_i(t) = L$ imply that $(s_1(t), ..., s_n(t), v_1(t), ..., v_n(t))$ is bounded on $[0, t_{\max})$. This implies $t_{\max} = +\infty$.

We show next by contradiction that $(s_1(t), ..., s_n(t), v_1(t), ..., v_n(t)) \in D$ for all $t \in [0, t_{\max})$. Therefore, we next assume that there exists $t \in [0, t_{\max})$ such that $(s_1(t), ..., s_n(t), v_1(t), ..., v_n(t)) \notin D$. By virtue of the fact that $(s_1(0), ..., s_n(0), v_1(0), ..., v_n(0)) \in D$, there exists a neighborhood of 0 such that $(s_1(t), ..., s_n(t), v_1(t), ..., v_n(t)) \in D$ for all $t \in [0, t_{\max})$ in this neighborhood (recall (3.7)). Consequently, if we define

$$T := \inf\left\{t \in [0, t_{\max}) : (s_1(t), ..., s_n(t), v_1(t), ..., v_n(t)) \notin D\right\} \qquad (7.12)$$

it follows that $T \in (0, t_{max})$. Notice that the case $(s_1(T),...,s_n(T),v_1(T),...,v_n(T)) \in D$ is excluded (since there would be a neighborhood of $T$ with $(s_1(t),...,s_n(t),v_1(t),...,v_n(t)) \in D$ for all $t \in [0, t_{max})$ in this neighborhood and that contradicts definition (7.12)). Notice also that definition (7.12) implies that

$$(s_1(t),...,s_n(t),v_1(t),...,v_n(t)) \in D \text{ for all } t \in [0,T) \tag{7.13}$$

and since $(s_1(T),...,s_n(T),v_1(T),...,v_n(T)) \notin D$, we conclude from (3.7) that (7.3) holds. Next, define the function $\Phi:[0,T) \to \Re$ by means of the formula (7.4). Notice that (7.3) and definition (7.4) imply that $\lim_{t \to T^-}(\Phi(t)) = +\infty$. Using (2.1), (2.2), (3.6) and definition (7.4) we obtain (7.5) for $t \in [0,T)$.

Using (2.1), (2.2), (3.6), (3.7), (3.1), (3.2), and the fact that $f, g, \kappa: \Re \to \Re_+$ are non-negative functions, we obtain inequalities (7.6). Moreover, (7.13) and definition (3.7) imply inequality (7.7). By virtue of (3.2) there exists $r > 0$ so that $v_{max} = k(\lambda - a - r)$. Using (7.6) and the fact that $v_0 \equiv v_n$, we get (7.8) for all $t \in [0,T)$, $i = 1,...,n$. Consequently, if $s_i(t) - a - k^{-1}(v_i(t) - v_{i-1}(t)) \geq r$ for certain $i = 1,...,n$ then we get from (7.8) the inequality $\dfrac{v_i(t) - v_{i-1}(t) + k^{-1}(\dot{v}_i(t) - \dot{v}_{i-1}(t))}{\left(s_i(t) - a - k^{-1}(v_i(t) - v_{i-1}(t))\right)^2} \leq \dfrac{v_{max}}{r^2}$.

On the other hand, if $s_i(t) - a - k^{-1}(v_i(t) - v_{i-1}(t)) < r$ then $s_i(t) < a + r + k^{-1}(v_i(t) - v_{i-1}(t))$, which combined with (7.13), (3.7) and the fact that $v_{max} = k(\lambda - a - r)$ gives $s(t) < a + r + k^{-1}v_{max} = \lambda$. Therefore, in this case we get from (7.6) and (2.1), (2.2), (3.6), (3.3):

$$v_i(t) - v_{i-1}(t) + k^{-1}(\dot{v}_i(t) - \dot{v}_{i-1}(t)) \leq v_i(t) + k^{-1}\dot{v}_i(t)$$
$$= k^{-1}f(s_i(t)) + k^{-1}g(s_i(t))v_{i-1}(t) = 0$$

Consequently, in any case, we obtain (7.9) for all $t \in [0,T)$, $i = 1,...,n$. Therefore, we obtain (7.10) from (7.4), (7.5), (7.6), (7.7) and (7.9) for all $t \in [0,T)$. The differential inequality (7.10) implies estimate (7.11). Estimate (7.11) contradicts the implication $\lim_{t \to T^-}(\Phi(t)) = +\infty$. The proof is complete. ◁

The proofs of Theorem 3 and Theorem 4 are exactly the same. They are provided next.

**Proof of Theorem 3 and Theorem 4:** Let an (arbitrary) input $v_0 \in C^1(\Re_+)$ that satisfies (3.5) and an (arbitrary) point $(s_{1,0},...,s_{n,0},v_{1,0},...,v_{n,0}) \in D(v_0(0))$ be given.

The fact that the initial-value problem (2.1), (2.2) with (4.6), initial condition $(s_1(0),...,s_n(0),v_1(0),...,v_n(0)) = (s_{1,0},...,s_{n,0},v_{1,0},...,v_{n,0})$ has a unique solution $(s_1(t),...,s_n(t),v_1(t),...,v_n(t))$ defined for all $t \geq 0$ that satisfies $(s_1(t),...,s_n(t),v_1(t),...,v_n(t)) \in D(v_0(t))$ for all $t \geq 0$, is a direct consequence of Theorem 1, properties (4.1), (4.2), (4.3), definition (4.5) and the fact that properties (3.1), (3.2), (3.3) hold for the functions $f(s) = (k - g(s))G(s)$, $\kappa(s) \equiv k$.

We define the following non-negative functions for $i = 1,...,n$

$$V_i(s_i, v_i) = \frac{1}{2}(v_i - v^*)^2 + \frac{1}{4q}(v_i - G(s_i))^2 + \int_{s^*}^{s_i} (k - g(z))(G(z) - v^*) dz \qquad (7.14)$$

where $q > 0$ is an arbitrary constant. By taking into account (4.1), (2.1), (2.2), (4.6), and by using the inequalities

$$2(v_i - v^*)(v_{i-1} - v^*) \leq (v_i - v^*)^2 + (v_{i-1} - v^*)^2,$$

$$|v_i - G(s_i)||v_{i-1} - v^*| \leq \frac{(v_i - G(s_i))^2}{2q} + q(v_{i-1} - v^*)^2,$$

we obtain for all $q > 0$, $i = 1, ..., n$

$$\begin{aligned}
\dot{V}_i &= -k(v_i - v^*)^2 - \frac{1}{2q}(k - g(s_i))(v_i - G(s_i))^2 \\
&+ \left(k(v_i - v^*) - (k - g(s_i))(v_i - G(s_i))\right)(v_{i-1} - v^*) \\
&\leq -\frac{k}{2}(v_i - v^*)^2 + \frac{1}{2}(k + q(k - g(s_i)))(v_{i-1} - v^*)^2
\end{aligned} \qquad (7.15)$$

Estimate (7.15) gives the following estimate for all $q > 0$, $i = 1, ..., n$

$$\dot{V}_i \leq -\frac{k}{2}(v_i - v^*)^2 + \frac{1}{2}(1 + q)k(v_{i-1} - v^*)^2$$

which after integration implies that for all $t \geq 0$, $q > 0$, $i = 1, ..., n$:

$$\int_0^t (v_i(\tau) - v^*)^2 d\tau + \frac{2}{k} V_i(t) \leq \frac{2}{k} V_i(0) + (1 + q) \int_0^t (v_{i-1}(\tau) - v^*)^2 d\tau \qquad (7.16)$$

Estimate (4.7) is a direct consequence of (7.16) and definition (7.14).

Next we use again (4.1), (2.1), (2.2), (4.6), (7.14) and the inequalities

$$2(v_i - v^*)(v_{i-1} - v^*) \leq (v_i - v^*)^2 + (v_{i-1} - v^*)^2,$$

$$|v_i - G(s_i)||v_{i-1} - v^*| \leq \frac{1}{4q}(v_i - G(s_i))^2 + q(v_{i-1} - v^*)^2,$$

$$|G(s_i) - v^*||v_i - v^*| \leq \frac{\varepsilon}{2}(G(s_i) - v^*)^2 + \frac{1}{2\varepsilon}(v_i - v^*)^2, \text{ for all } \varepsilon > 0$$

to get the following estimate for all $\varepsilon > 0$:

$$2\dot{V}_i \leq -k(v_i - v^*)^2 - \frac{1}{2q}(k - g_{\max})(v_i - v^*)^2$$
$$-\frac{1}{2q}(k - g_{\max})(G(s_i) - v^*)^2 + \frac{1}{2\varepsilon q}(k - g_{\max})(v_i - v^*)^2 \qquad (7.17)$$
$$+\frac{\varepsilon}{2q}(k - g_{\max})(G(s_i) - v^*)^2 + (1 + 2q)k(v_{i-1} - v^*)^2$$

Setting $\varepsilon = \dfrac{k - g_{\max}}{2qk + k - g_{\max}}$, it follows from (7.17) that

$$\dot{V}_i \leq -\frac{k(k - g_{\max})}{4qk + 2(k - g_{\max})}(G(s_i) - v^*)^2 + \frac{k}{2}(1 + 2q)(v_{i-1} - v^*)^2 \qquad (7.18)$$

Inequality (7.18) and the fact that $V_i(t) \geq 0$ (recall (7.14)) imply that the following estimate holds for all $t \geq 0$:

$$\int_0^t (G(s_i(\tau)) - v^*)^2 d\tau \leq 2\frac{(1 + 2q)k - g_{\max}}{k(k - g_{\max})} V_i(0)$$
$$+ (1 + 2q)\frac{(1 + 2q)k - g_{\max}}{k - g_{\max}} \int_0^t (v_{i-1}(\tau) - v^*)^2 d\tau \qquad (7.19)$$

Estimate (4.8) is a direct consequence of (7.19) and definition (7.14).

We next define for $i = 1, ..., n$:
$$w_i := v_i - G(s_i) \qquad (7.20)$$

Due to (2.1), (2.2), (4.6) and (7.20) we have for all $t \geq 0$, $i = 1, ..., n$:
$$\dot{w}_i = -(k - g(s_i))w_i \qquad (7.21)$$

The solution of (7.21) is given by the formula
$$w_i(t) = w_i(0) \exp\left(-kt + \int_0^t g(s_i(\tau))d\tau\right), \text{ for all } t \geq 0, \ i = 1, ..., n \qquad (7.22)$$

Estimate (4.10) is a direct consequence of (4.1), (7.22) and definition (7.20).

Next, notice next that due to (2.1), (2.2), (4.6) and definition (7.21), we have for all $t \geq 0$, $i = 1, ..., n$:
$$\dot{v}_i = -(k - g(s_i))w + g(s_i)(v_{i-1} - v^*) - g(s_i)(v_i - v^*) \qquad (7.23)$$

It follows from (7.22) and (7.23) that

$$v_i(t) - v^* = (v_i(0) - v^*) \exp\left(-\int_0^t g(s_i(\tau)) d\tau\right)$$

$$+ \int_0^t g(s_i(l)) \exp\left(-\int_l^t g(s_i(\tau)) d\tau\right) (v_{i-1}(l) - v^*) dl \qquad (7.24)$$

$$- w_i(0) \exp\left(-\int_0^t g(s_i(\tau)) d\tau\right) \int_0^t (k - g(s_i(l))) \exp(-kl) dl$$

Finally, using (7.20), (7.24), the triangle inequality and the fact that $g$ is a non-negative function, we obtain

$$|v_i(t) - v^*| \leq |v_i(0) - v^*| + k|w_i(0)| \int_0^t \exp(-kl) dl$$

$$+ \int_0^t g(s_i(l)) \exp\left(-\int_l^t g(s_i(\tau)) d\tau\right) dl \sup_{0 \leq l \leq t} \left(|v_{i-1}(l) - v^*|\right)$$

$$\leq |v_i(0) - v^*| + |w_i(0)| \qquad (7.25)$$

$$+ \exp\left(-\int_0^t g(s_i(\tau)) d\tau\right) \int_0^t \frac{d}{dl}\left(\exp\left(\int_0^l g(s_i(\tau)) d\tau\right)\right) dl \sup_{0 \leq l \leq t}\left(|v_{i-1}(l) - v^*|\right)$$

$$\leq 2|v_i(0) - v^*| + |G(s_i(0)) - v^*| + \sup_{0 \leq l \leq t}\left(|v_{i-1}(l) - v^*|\right)$$

Estimate (4.9) is a direct consequence of estimate (7.25). The proof is complete. ◁

Next we provide the proof of Theorem 5. The proof of Theorem 5 is performed by constructing a Lyapunov function for system (2.1), (2.2), (4.6) with $v_0 = v^*$.

**Proof of Theorem 5:** By virtue of Theorem 3, the set $D(v^*) \subset \Re^{2n}$ is positively invariant for system (2.1), (2.2), (4.6) with $v_0 = v^*$. Therefore, Proposition 1.4.5 on page 20 in [2] guarantees that the set $\overline{D(v^*)} \subset \Re^{2n}$ is positively invariant for system (2.1), (2.2), (4.6) with $v_0 = v^*$.
We next show that the equilibrium point $(s^*,...,s^*,v^*,...,v^*) \in \overline{D(v^*)}$ for system (2.1), (2.2), (4.6) with $v_0 = v^*$ defined on $\overline{D(v^*)}$ is Globally Asymptotically Stable. For arbitrary constant $c > 0$, we define the family of functions

$$V_i(s_i, v_i) = \frac{1}{2}(v_i - v^*)^2 + \frac{c}{2}(v_i - G(s_i))^2 + \int_{s^*}^{s_i} (k - g(z))(G(z) - v^*) dz, \quad i = 1,...,n \qquad (7.26)$$

Using definition (7.26), the fact that $g$ is non-negative and the inequalities

$$|v_i - G(s_i)||v_{i-1} - v^*| \leq \frac{c}{2}(v_i - G(s_i))^2 + \frac{1}{2c}(v_{i-1} - v^*)^2$$

$$2(v_i - v^*)(v_{i-1} - v^*) \leq (v_i - v^*)^2 + (v_{i-1} - v^*)^2$$

we obtain for $i = 1,...,n$:

$$\dot{V}_i = -k\left(v_i - v^*\right)^2 - c(k - g(s_i))\left(v_i - G(s_i)\right)^2$$
$$-(k - g(s_i))\left(v_i - G(s_i)\right)(v_{i-1} - v^*) + k\left(v_i - v^*\right)\left(v_{i-1} - v^*\right) \quad (7.27)$$
$$\leq -\frac{k}{2}\left(v_i - v^*\right)^2 - \frac{c}{2}(k - g(s_i))\left(v_i - G(s_i)\right)^2 + \frac{k}{2}\left(\frac{1}{c} + 1\right)\left(v_{i-1} - v^*\right)^2$$

Define the coefficients $Q_i \geq 1$ for $i = 1,...,n$ by means of the equations:

$$Q_i = 1 + Q_{i+1}\left(\frac{1}{c} + 1\right) \text{ for } i = 1,...,n-1 \text{ and } Q_n = 1 \quad (7.28)$$

Moreover, define the Lyapunov function:

$$V(s_1,...,s_n,v_1,...,v_n) = \sum_{i=1}^{n} Q_i V_i(s_i, v_i) \quad (7.29)$$

Due to (4.1), (7.27), (7.28), (7.29) and the fact that $v_0 = v^*$, we get:

$$\dot{V} \leq -\frac{k}{2}\sum_{i=1}^{n}\left(v_i - v^*\right)^2 - \frac{c}{2}(k - g_{\max})\sum_{i=1}^{n} Q_i\left(v_i - G(s_i)\right)^2 \quad (7.30)$$

Notice that due to the fact that the function $G$ defined by (4.5) is strictly increasing on $[\lambda, +\infty)$, we can conclude that the right hand side of (7.30) is negative for all $(s_1,...,s_n,v_1,...,v_n) \in \overline{D(v^*)}$ with $(s_1,...,s_n,v_1,...,v_n) \neq (s^*,...,s^*,v^*,...,v^*)$. Again, due to the fact that the function $G$ defined by (4.5) is strictly increasing on $[\lambda, +\infty)$, definitions (7.26), (7.29) and equation (4.4) guarantee that $V$ is positive for all $(s_1,...,s_n,v_1,...,v_n) \in \overline{D(v^*)}$ with $(s_1,...,s_n,v_1,...,v_n) \neq (s^*,...,s^*,v^*,...,v^*)$. Therefore, Theorem 2.13 on page 73 in [7] shows that it suffices to show that the function $V$ defined by (7.29) is uniformly unbounded (see Definition 2.8 on page 70 in [7]). In other words, it suffices to show that for every $M > 0$ the sublevel set $\left\{(s_1,...,s_n,v_1,...,v_n) \in \overline{D(v^*)} : V(s_1,...,s_n,v_1,...,v_n) \leq M\right\}$ is bounded. The fact that the function $G$ defined by (4.5) is strictly increasing on $[\lambda, +\infty)$ with $G(s) \equiv 0$ for $s < \lambda$ in conjunction with (4.1) implies that

$$\int_{s^*}^{s}(k - g(z))\left(G(z) - v^*\right)dz \geq (k - g_{\max})\int_{s^*}^{s}\left(G(z) - v^*\right)dz$$
$$\geq (k - g_{\max})\int_{s^*+1}^{s}\left(G(z) - v^*\right)dz \geq (k - g_{\max})\left(G(s^* + 1) - v^*\right)\left(s - s^* - 1\right)$$

for all $s \geq s^* + 1$. Consequently, the inequality $V(s_1,...,s_n,v_1,...,v_n) \leq M$ in conjunction with the fact that $Q_i \geq 1$ for $i = 1,...,n$, implies that $s_i \leq s^* + 1 + \dfrac{M}{(k - g_{\max})\left(G(s^* + 1) - v^*\right)}$ for $i = 1,...,n$. Thus, it follows that the sublevel set $\left\{(s_1,...,s_n,v_1,...,v_n) \in \overline{D(v^*)} : V(s_1,...,s_n,v_1,...,v_n) \leq M\right\}$ is bounded for every $M > 0$.

Next we assume that $g$ is of class $C^1$ in a neighborhood of $s^* > \lambda$. It follows that there is a neighborhood of the equilibrium point $(s^*,...,s^*,v^*,...,v^*)$ for which the right hand side of system (2.1), (2.2), (4.6) with $v_0 = v^*$ is continuously differentiable. Therefore, by virtue of Theorem 3.7 on page 127 in [15] in order to show that the equilibrium point $(s^*,...,s^*,v^*,...,v^*)$ it suffices to show that the Jacobian matrix at $(s^*,...,s^*,v^*,...,v^*)$ is a Hurwitz matrix. The Jacobian matrix at $(s^*,...,s^*,v^*,...,v^*)$ has the following lower diagonal block structure

$$\begin{pmatrix} A & 0 & \cdots & 0 \\ * & A & \ldots & 0 \\ \vdots & \vdots & & \vdots \\ * & * & \ldots & A \end{pmatrix}$$

where

$$A = \begin{pmatrix} 0 & -1 \\ (k-g(s^*))g(s^*) & -k \end{pmatrix}$$

Therefore the Jacobian matrix at $(s^*,...,s^*,v^*,...,v^*)$ has two eigenvalues $\lambda_1 = -g(s^*)$ and $\lambda_2 = -(k-g(s^*))$, each one with algebraic multiplicity $n$. Both eigenvalues are negative and consequently, the Jacobian matrix at $(s^*,...,s^*,v^*,...,v^*)$ is a Hurwitz matrix.
The proof is complete.  ◁

**Proof of Theorem 6:** By virtue of Theorem 4, the set $\tilde{D} = D \cap \left\{ (s_1,...,s_n,v_1,...,v_n) \in \Re^{2n} : \sum_{i=1}^{n} s_i = L \right\}$ is positively invariant for system (2.1), (2.2), (4.6) with $v_0 = v_n$. Therefore, Proposition 1.4.5 on page 20 in [2] guarantees that the set $\Omega \subset \Re^{2n}$ defined by (5.3), is positively invariant for system (2.1), (2.2), (4.6) with $v_0 = v_n$.

Notice that since $s^* = L/n$, we have that $\sum_{i=1}^{n}(s_i - s^*) = 0$ and by using condition (5.2) with $x_i = s_i - s^*$, $i = 1,...,n$ it holds that

$$(s_1 - s_n)^2 + \sum_{i=2}^{n}(s_i - s_{i-1})^2 \geq \mu_n \sum_{i=1}^{n}(s_i - s^*)^2 \tag{7.31}$$

with $\mu_n > 0$ as defined in (5.1). Next consider a constant $c > 0$ so that

$$\frac{p}{2}\mu_n > 2M + \frac{2}{c(k-g_{\max})} \tag{7.32}$$

which is feasible since $M$ satisfies the strict inequality $M < \frac{p\mu_n}{4}$. Define the Lyapunov function

$$V(s_1,...,s_n,v_1,...,v_n) = \frac{1}{2}\sum_{i=1}^{n}\left(s_i - s^*\right)^2 + \frac{c}{2}\sum_{i=1}^{n}\left(v_i - G(s_i)\right)^2 \tag{7.33}$$

with $s_0 = s_n$. Using (7.33), (2.1), (2.2), (4.6) with $v_0 = v_n$, the time-derivative of $V$ can be calculated as follows:

$$\dot{V} = \sum_{i=1}^{n}(s_i - s^*)(v_{i-1} - v_i) - c\sum_{i=1}^{n}(k - g(s_i))(v_i - G(s_i))^2$$

$$= \sum_{i=1}^{n}(s_i - s^*)(G(s_{i-1}) - G(s_i)) + \sum_{i=1}^{n}(s_i - s^*)(v_{i-1} - G(s_{i-1})) \quad (7.34)$$

$$- \sum_{i=1}^{n}(s_i - s^*)(v_i - G(s_i)) - c\sum_{i=1}^{n}(k - g(s_i))(v_i - G(s_i))^2$$

Notice that since $(s_1,...,s_n,v_1,...,v_n) \in \Omega$ it follows from (5.3) and (3.7) that $s_i \in [a, L-(n-1)a]$ for $i = 1,...,n$. Using (4.1), the fact that $v_0 = v_n$ and completing the squares in (7.34), we get

$$\dot{V} \leq \sum_{i=1}^{n}(s_i - s^*)(G(s_{i-1}) - G(s_i))$$

$$+ \frac{2}{c(k - g_{max})}\sum_{i=1}^{n}(s_i - s^*)^2 + \frac{c(k - g_{max})}{4}\sum_{i=1}^{n}(v_i - G(s_i))^2$$

$$+ \frac{c(k - g_{max})}{4}\sum_{i=0}^{n-1}(v_i - G(s_i))^2 - c\sum_{i=1}^{n}(k - g(s_i))(v_i - G(s_i))^2 \quad (7.35)$$

$$\leq \sum_{i=1}^{n}(s_i - s^*)(G(s_{i-1}) - G(s_i))$$

$$+ \frac{2}{c(k - g_{max})}\sum_{i=1}^{n}(s_i - s^*)^2 - \frac{c(k - g_{max})}{2}\sum_{i=1}^{n}(v_i - G(s_i))^2$$

By adding and subtracting terms in (7.35) we obtain the following inequality:

$$\dot{V} \leq p\sum_{i=1}^{n}(s_i - s^*)(s_{i-1} - s_i)$$

$$+ \sum_{i=1}^{n}(s_i - s^*)\left(G(s_{i-1}) - p(s_{i-1} - s^*) - G(s_i) + p(s_i - s^*)\right) \quad (7.36)$$

$$+ \frac{2}{c(k - g_{max})}\sum_{i=1}^{n}(s_i - s^*)^2 - \frac{c(k - g_{max})}{2}\sum_{i=1}^{n}(v_i - G(s_i))^2$$

Notice next that by using the facts that $\sum_{i=1}^{n}s_i = L$ and $s_0 = s_n$, we get

$$\sum_{i=1}^{n}(s_i - s^*)(s_{i-1} - s_i) = \sum_{i=1}^{n} s_i(s_{i-1} - s_i) - s^* \sum_{i=1}^{n} s_{i-1} + s^* \sum_{i=1}^{n} s_i$$

$$= \sum_{i=1}^{n} s_i(s_{i-1} - s_i) - s^*L + s^*L = -\sum_{i=1}^{n} s_i(s_i - s_{i-1})$$

$$= -\sum_{i=1}^{n}(s_i - s_{i-1})^2 - \sum_{i=1}^{n} s_{i-1}(s_i - s_{i-1})$$

$$= -\frac{1}{2}\sum_{i=1}^{n}(s_i - s_{i-1})^2 - \sum_{i=1}^{n} s_{i-1}(s_i - s_{i-1}) - \frac{1}{2}\sum_{i=1}^{n}(s_i - s_{i-1})^2 \quad (7.37)$$

$$= -\frac{1}{2}\sum_{i=1}^{n}(s_i - s_{i-1})^2 - \sum_{i=1}^{n} s_{i-1}s_i + \sum_{i=1}^{n} s_{i-1}^2 - \frac{1}{2}\sum_{i=1}^{n} s_i^2 - \frac{1}{2}\sum_{i=1}^{n} s_{i-1}^2 + \sum_{i=1}^{n} s_{i-1}s_i$$

$$= -\frac{1}{2}\sum_{i=1}^{n}(s_i - s_{i-1})^2$$

Hence, it follows from (7.36) and (7.37) that

$$\dot{V} \leq -\frac{p}{2}\sum_{i=1}^{n}(s_i - s_{i-1})^2$$

$$+ \sum_{i=1}^{n}(s_i - s^*)\big(G(s_{i-1}) - p(s_{i-1} - s^*) - G(s_i) + p(s_i - s^*)\big) \quad (7.38)$$

$$+ \frac{2}{c(k - g_{\max})}\sum_{i=1}^{n}(s_i - s^*)^2 - \frac{c(k - g_{\max})}{2}\sum_{i=1}^{n}(v_i - G(s_i))^2$$

Notice now that due to the condition $s_0 = s_n$ and (7.31) the following inequality holds:

$$\sum_{i=1}^{n}(s_i - s_{i-1})^2 \geq \mu_n \sum_{i=1}^{n}(s_i - s^*)^2 \quad (7.39)$$

From (7.38), (7.39) and inequality (5.4) we obtain the following estimate

$$\dot{V} \leq -\frac{p}{2}\mu_n \sum_{i=1}^{n}(s_i - s^*)^2$$

$$+ \sum_{i=1}^{n}|s_i - s^*|\big|G(s_{i-1}) - v^* - p(s_{i-1} - s^*)\big| + \sum_{i=1}^{n}|s_i - s^*|\big|G(s_i) - v^* - p(s_i - s^*)\big|$$

$$+ \frac{2}{c(k - g_{\max})}\sum_{i=1}^{n}(s_i - s^*)^2 - \frac{c(k - g_{\max})}{2}\sum_{i=1}^{n}(v_i - G(s_i))^2 \quad (7.40)$$

$$- \left(\frac{p}{2}\mu_n - M - \frac{2}{c(k - g_{\max})}\right)\sum_{i=1}^{n}(s_i - s^*)^2$$

$$+ M\sum_{i=1}^{n}|s_i - s^*||s_{i-1} - s^*| - \frac{c(k - g_{\max})}{2}\sum_{i=1}^{n}(v_i - G(s_i))^2$$

$$\leq -\left(\frac{p}{2}\mu_n - 2M - \frac{2}{c(k - g_{\max})}\right)\sum_{i=1}^{n}(s_i - s^*)^2 - \frac{c(k - g_{\max})}{2}\sum_{i=1}^{n}(v_i - G(s_i))^2$$

By virtue of (7.40), (7.32) and (7.33) there exists a constant $\varphi > 0$ such that

$$\dot{V} \leq -2\varphi V \qquad (7.41)$$

On the other hand, inequality (4.1) and definitions (4.4), (4.5) imply the inequality $|G(s) - v^*| \leq g_{max}|s - s^*|$ for all $s \geq a$. Consequently, we get from definition (7.33) that

$$V(s_1,...,s_n,v_1,...,v_n) \leq \left(\frac{1}{2} + c + cg_{max}^2\right)\left|(s_1 - s^*,...,s_n - s^*, v_1 - v^*,...,v_n - v^*)\right|^2 \qquad (7.42)$$

Let an (arbitrary) point $(s_{1,0},...,s_{n,0},v_{1,0},...,v_{n,0}) \in \Omega$ be given. Since the set $\Omega \subset \Re^{2n}$ defined by (5.3), is positively invariant for system (2.1), (2.2), (4.6) with $v_0 = v_n$, it follows that the initial-value problem (2.1), (2.2), (4.6) with $v_0 = v_n$, initial condition $(s_1(0),...,s_n(0),v_1(0),...,v_n(0)) = (s_{1,0},...,s_{n,0},v_{1,0},...,v_{n,0})$ has a unique solution $(s_1(t),...,s_n(t),v_1(t),...,v_n(t))$ defined for all $t \geq 0$ that satisfies $(s_1(t),...,s_n(t),v_1(t),...,v_n(t)) \in \Omega$ for all $t \geq 0$. By integrating (7.41), we obtain for all $t \geq 0$:

$$V(s_1(t),...,s_n(t),v_1(t),...,v_n(t)) \leq \exp(-2\varphi t)V(s_1(0),...,s_n(0),v_1(0),...,v_n(0)) \qquad (7.43)$$

Inequality (7.43) in conjunction with inequality (7.42) and definition (7.33) gives the following estimate for all $t \geq 0$:

$$\sum_{i=1}^{n}\left(s_i(t) - s^*\right)^2 + c\sum_{i=1}^{n}\left(v_i(t) - G(s_i(t))\right)^2$$
$$\leq \exp(-2\varphi t)\left(1 + 2c + 2cg_{max}^2\right)\left|(s_1(0) - s^*,...,s_n(0) - s^*, v_1(0) - v^*,...,v_n(0) - v^*)\right|^2 \qquad (7.44)$$

Using the inequality $|G(s) - v^*| \leq g_{max}|s - s^*|$ for all $s \geq a$, we obtain from (7.44) for all $t \geq 0$:

$$\left|(s_1(t) - s^*,...,s_n(t) - s^*, v_1(t) - v^*,...,v_n(t) - v^*)\right|^2 =$$
$$= \sum_{i=1}^{n}\left(s_i(t) - s^*\right)^2 + \sum_{i=1}^{n}\left(v_i(t) - v^*\right)^2$$
$$\leq \sum_{i=1}^{n}\left(s_i(t) - s^*\right)^2 + 2\sum_{i=1}^{n}\left(v_i(t) - G(s_i(t))\right)^2 + 2\sum_{i=1}^{n}\left(v^* - G(s_i(t))\right)^2$$
$$\leq 2\sum_{i=1}^{n}\left(v_i(t) - G(s_i(t))\right)^2 + \left(1 + 2g_{max}^2\right)\sum_{i=1}^{n}\left(s_i(t) - s^*\right)^2$$
$$\leq \left(\frac{2}{c} + 1 + 2g_{max}^2\right)\left(\sum_{i=1}^{n}\left(s_i(t) - s^*\right)^2 + c\sum_{i=1}^{n}\left(v_i(t) - G(s_i(t))\right)^2\right)$$
$$\leq \exp(-2\varphi t)R^2\left|(s_1(0) - s^*,...,s_n(0) - s^*, v_1(0) - v^*,...,v_n(0) - v^*)\right|^2$$

where $R^2 := \left(\dfrac{2}{c} + 1 + 2g_{max}^2\right)\left(1 + 2c + 2cg_{max}^2\right)$. The above estimate implies the following inequality for all $t \geq 0$:

$$\begin{aligned}&\left|\left(s_1(t) - s^*, \ldots, s_n(t) - s^*, v_1(t) - v^*, \ldots, v_n(t) - v^*\right)\right| \\ &\leq \exp(-\varphi t) R \left|\left(s_1(0) - s^*, \ldots, s_n(0) - s^*, v_1(0) - v^*, \ldots, v_n(0) - v^*\right)\right|\end{aligned} \quad (7.45)$$

which directly proves global exponential stability. The proof is complete. ◁

## 8. Concluding Remarks

The present work proposed a novel nonlinear adaptive cruise controller for vehicular platoons operating on an open road or a ring-road. The proposed controller is a nonlinear function of the distance between successive vehicles and their speed. Certain conditions were derived that guarantee safety in terms of collision avoidance and bounded vehicle speeds by explicitly characterizing a set of admissible initial conditions and the set of allowable inputs. It is shown that a platoon of vehicles with this controller is $L_p$ string stable, and all vehicles will converge to the desired speed/spacing configuration from any initial condition. Future work will address the impact of sensor and actuator delays, as well as the effects of nudging on the stability, string stability and safety of vehicular platoons.

## Acknowledgments

The research leading to these results has received funding from the European Research Council under the European Union's Horizon 2020 Research and Innovation programme/ ERC Grant Agreement n. [833915], project TrafficFluid.